\documentclass[a4paper]{article}

\usepackage[margin=1in]{geometry} 

\usepackage{amsmath}
\usepackage{amsthm}
\usepackage{amssymb}

\usepackage[utf8]{inputenc}

\usepackage{url}
\usepackage{multirow}
\usepackage{mathtools}
\usepackage{xspace}
\usepackage{hyperref}
\usepackage{dirtytalk}
\usepackage{graphicx}

\usepackage{tabularx}
\usepackage{booktabs}
\usepackage{subcaption}
\usepackage{caption}

\usepackage{tikz}
\usepackage{pgfplots}
\usetikzlibrary{positioning}
\usetikzlibrary{pgfplots.groupplots}

\newcommand*{\eg}{e.g.\@\xspace}
\newcommand*{\ie}{i.e.\@\xspace}

\newtheorem{definition}{Definition}

\title{Towards Standardized Mobility Reports with User-Level Privacy}
\author{Alexandra Kapp$^1$
	\and 
	Saskia Nu\~nez von Voigt$^2$
	\and Helena Mihaljevi\'{c}$^3$
	\and Florian Tschorsch$^4$
}

\date{\small{
	$^1$ORCiD:  0000-0002-8348-8958, Hochschule für Technik und Wirtschaft Berlin, University of 
	Applied Sciences, 
	\texttt{kapp@htw-berlin.de}\\
	$^2$ Distributed Security Infrastructures Technische Universit\"at\\
	$^3$ORCiD: 0000-0003-0782-5382, Hochschule für Technik und Wirtschaft Berlin, University of 
	Applied Sciences\\
	$^4$ORCiD: 0000-0001-6716-7225, Distributed Security Infrastructures Technische Universit\"at
	Berlin	
}}

\begin{document}

\vspace{5cm}

\maketitle

\vspace{5cm}

\begin{abstract}
The importance of human mobility analyses is growing in both research and practice, 
especially as applications for urban planning and mobility rely on them. 
Aggregate statistics and visualizations play an essential role as building blocks of data explorations and 
summary reports, the latter being increasingly released to third parties such as municipal administrations 
or in the context of citizen participation.
However, such explorations already pose a threat to privacy as they 
reveal potentially sensitive location information, and thus should not be shared without further privacy 
measures. 

There is a substantial gap between state-of-the-art research on
privacy methods and their utilization in practice.
We thus conceptualize a standardized mobility report with differential privacy guarantees and implement it 
as open-source software to enable a privacy-preserving exploration 
of key aspects of mobility data in an easily accessible way. Moreover, we evaluate the benefits of limiting 
user contributions using three data sets relevant to research and practice. 
Our results show that even a strong limit on user contribution alters the original 
geospatial distribution only within a comparatively small range, while significantly reducing 
the error introduced by adding noise to achieve privacy guarantees.

\end{abstract}

\newpage

\section{Introduction}
\label{sec:intro}

Mobility data is becoming increasingly important - 
for urban planning~\cite{zhou_understanding_2018}, traffic 
management~\cite{naboulsi_large-scale_2016}, and smart city 
applications~\cite{creutzig_smart_2021}. To tackle challenges posed by the 
climate crisis,
cities require movement data to address smart traffic management 
\cite{ravish_intelligent_2021, djahel_communications-oriented_2015}, 
shift car use to climate-friendlier options~\cite{nieuwenhuijsen_car_2016}, 
offer broader access to public transport~\cite{faivre_darcier_measuring_2014}, 
and steer ever-growing mobility options such as e-scooters 
\cite{creutzig_smart_2021}, to name a few. The current pandemic 
further highlights the usefulness of such data as these form 
the basis for analyses~\cite{gao_mapping_2020,aktay_google_2020} or 
simulations of pandemic progression~\cite{pesavento_data-driven_2020}.

At the same time, mobility data comprises high-dimensional, sensitive information and thus poses 
challenges towards ensuring privacy.
For example, just a few recorded locations are sufficient to
re-identify an individual~\cite{de_montjoye_unique_2013}, potentially revealing private information such as 
home and work locations, political interests or religious beliefs. Recent discussions on storage and 
processing of mobility data for pandemic management highlight that many citizens are willing to disclose 
their movement data only under extensive privacy 
guarantees~\cite{hargittai_americans_2020, 
altmann_acceptability_2020}. Uncertainty about how to adequately protect user privacy and 
thus comply with applicable data protection laws, for example, makes many mobility providers reluctant to 
share their usage data with cities or otherwise release it.

Transparency of policy making can therefore only be guaranteed to a limited extent.
The Open Mobility Foundation facilitates data exchange between mobility providers and cities, thereby also 
striving to foster transparency of policy processes to the public. At the same time, they address the 
inherent privacy issues of the release of raw mobility data and suggest data minimization, aggregations 
and reports as potential 
countermeasures~\cite{open_mobility_foundation_practical_2020}. For a lot of 
decisions reporting aggregate statistics is sufficient as many use cases need no fine granular trajectory 
data. In Germany, for example, according to road traffic regulations a street can be redesignated as a 
bicycle street if a high volume of bicycle traffic can be proven. 
Similarly, aggregated mobility data is sufficient to monitor the impact of 
measures like new bike infrastructure or shared-mobility docking stations on usage behavior. Beyond 
monitoring tasks, aggregated statistics are highly valuable for more complex 
use cases, such as traffic 
models~\cite{ziemke_matsim_2019} or identification of
optimal e-scooter parking zones~\cite{zakhem_micromobility_2021}, since these 
require exploratory data 
analyses as a first step to familiarize with the data and
evaluate its suitability for the aspired use case. While aggregations provide some privacy they do 
not come with privacy \textit{guarantees}; in fact there are successful attacks on aggregated mobility data 
\cite{10.1145/3038912.3052620} that recover entire trajectories of individuals.

We propose a standardized mobility report comprising most relevant and frequently applied mobility 
measures that is equipped with privacy guarantees. For this purpose, we draw on extensive research on 
the analysis of movement data that requires the simultaneous examination of spatial and temporal 
properties
together with the moving object itself~\cite{andrienko_understanding_2016}. 
Depending on the data source, mobility data can differ greatly in its format and structure, spatial and 
temporal granularity, 
or content. Thus, it is hardly possible to comprise a standardized mobility report that suits all data sources 
and use cases. We carefully aimed to maintain the report as general as possible while knowing that it 
cannot fit all needs. Specifically, our report provides high level key insights to mobility data sets that could 
be used to provide information to public administrations, to the public, or for company's internal knowledge 
sharing. The report is especially suited for data containing single trips with an origin and destination (e.g., 
bike-sharing transaction history) and use cases about mobility on an aggregated 
level (e.g., usage of a new introduced bike-sharing service). It is also aimed at the analysis of staypoints 
and does not 
cover route information such as traffic volumes or the average speed on route segments.
Since mobility data includes sensitive information, we additionally see the need for privacy protection and 
apply methods for securing differential privacy~\cite{dwork_calibrating_2006} 
which is considered future-proof,
e.g. the privacy guarantees hold regardless of the attacker knowledge or
additional information (that becomes available later).

Unlike an iterative approach of arbitrary exploratory analyses, a standardized report allows for optimizing 
the allocation of a predefined privacy budget, as exploratory analyses might otherwise come at a high 
privacy cost~\cite{nunez_von_voigt_every_2020}. In order to keep the privacy costs as low as possible 
while ensuring high usefulness of the report, both the selection and detailed design of the analyses were 
optimized.

While differential privacy is well understood 
for aggregations of general tabular data, there is little research on how to 
apply such a mechanism to aggregations of mobility data.
Due to the nature of mobility data, differential privacy is difficult to apply as mobility traces of people are 
inherently individual with arbitrarily long sequences and spatial and temporal dimensions covering an 
infinite range of values. Broadly speaking, differential 
privacy guarantees that an adversary can only identify the membership of an individual within a data set at 
a 
defined probability level. The lower the identification probability shall be maintained the more noise has to 
be added. Generally, the smaller the number of buckets of aggregations the lower the share of 
added noise: consider adding noise of +/- 5 items to one bucket with a count of 100 or to 10 buckets each 
with a count of 10. In the first case, the noisy number of items will differ by 
at most 5\%, while in the second 
case, the difference per bucket will be up to 50\%. The number of buckets of mobility data can be adjusted 
at the spatial dimension (e.g., 
aggregation on a 100 m grid vs. 1 km grid) and the temporal dimension (e.g., hourly aggregation vs. weekly 
aggregation). Also, the space of possible sequences can be adjusted, e.g., by only considering start and 
end connections instead of entire routes. However, the number of buckets increases drastically once 
certain 
attributes are cross tabulated, for example, looking at spatial distributions per hour of day, or investigating 
the relation of origins and destinations. Data sparsity is an additional issue, as the records are usually not 
equally distributed over all buckets. For example, a city center usually has high visit counts while other 
areas further outside only get a few visits. Thus, differentially private values are only reliable for certain 
buckets while others mainly consist of noise.

We aim to gain insights how differential privacy can be applied to aggregations of mobility data and what 
level of utility could be expected when applied to typical, real-world data sets. We therefore evaluate key 
measures of the report under different privacy 
scenarios on three data sets that are used extensively, especially in mobility 
research, e.g., to infer transportation modes~\cite{zheng_understanding_2008}, 
mine locations of interest 
\cite{zheng_mining_2009}, or to investigate the influence of parking-regulations on car-ownership 
\cite{gonzalez_what_2021}.
This provides an important orientation for the trade-off between privacy and utility of real mobility data 
sets. 
Moreover, a typical mobility data set can contain an arbitrary number of records per user; an upper bound 
of user contribution is typically not set or known upfront. Intuitively, the more records a user contributes to 
a data set the more likely they will be identified. Thus, to provide differential privacy at a user level, more 
noise needs to be added according to the potential contribution of a single user.
In order to achieve acceptable estimates of the sensitivity of the respective aggregation functions when 
applying noise to ensure user-level privacy, 
it is useful to limit the records per user 
~\cite{amin_bounding_2019,epasto_smoothly_2020,liu_learning_2020,wilson_differentially_2020}. 
However, it is not clear how to choose such an upper bound.
We thus analyze the trade-off 
between privacy and utility under user-level privacy guarantees 
and different choices of the limit of user contribution.

A standardized report for mobility data with differential privacy guarantees 
has the potential to simplify and accelerate secure analysis and releases of 
movement data.
Analyses of mobility data require specific geospatial skill sets and resources.
With regard to privacy it should be noted that large 
companies such as Google, Apple, or Microsoft already make use of differential privacy 
\cite{cormode_privacy_2018}, while smaller companies and the public sector often lack 
knowledge and resources required to keep up with big tech 
players~\cite{hopkins_machine_2021}. To 
facilitate access and increase usability, we thus provide an implementation of our differentially private 
mobility report as open-source Python code.

\textbf{Our Contributions.}
After discussing  research on mobility data analyses and
appropriate measures for guaranteeing differential privacy in
Section~\ref{sec:relatedWork},
we identify elementary measures commonly applied to human mobility
data (Section~\ref{sec:standardReport}) and compile a standardized mobility report.
We provide a differentially private version of the report and explain the resulting implications
(Section~\ref{sec:dpReport}).
We analyze the effect of user-level privacy and the interplay with
different bounds of user contribution on three data sets
(Section~\ref{sec:eval}), and
provide practical implications of a user-level differentially private mobility
report (Section~\ref{sec:discussion}).
In this way, we support data analysts in understanding the implications of a 
differentially private mobility report.

\section{Related Work}
\label{sec:relatedWork}

Data analysts use exploratory data analyses (EDA) to become familiar with data sets,
usually focusing on visual inspection. They aim to understand
the data, check its validity, detect outliers, identify possible patterns and
assess its suitability for further analyses or models
\cite{behrens_principles_1997,andrienko_spatial_2011}.
The concept of EDA was established by Tukey, who, for example, promoted the use of
the five-number summary to describe the distribution of numerical data
\cite{tukey_exploratory_1977}. The issue of standardizing EDA has already been addressed in literature, 
e.g.,~\cite{zuur_protocol_2010}. In addition, implementations of standardized 
data explorations have recently been made available, e.g., in form of the Python 
package
\texttt{pandas\_profiling}~\cite{pandasprofiling2019},
making exploratory analyses accessible for a broader group of practitioners.

The analysis of movement data poses
additional challenges, as it requires the simultaneous examination of spatial 
and temporal properties 
together with the moving object itself~\cite{andrienko_understanding_2016}. 
Therefore, mobility data need
specific analyses in addition to standard EDA methods that can capture and
convey their information.
Andrienko et al.~\cite{andrienko_visual_2013} have massively
contributed to visual analytics of different types of movement data that can be considered valuable 
for the creation of a mobility data report. However, no attempt has been made 
to standardize a set of analyses for a particular type of movement data and analytical context
such as human mobility in cities. 
Graser~\cite{graser_exploratory_2021} argues that a standardized protocol for
an EDA can hardly cover all
kinds of movement data at once and 
presents a protocol for an EDA with the goal to identify problems in GPS data, 
such as unrealistic jumps in individual trajectories. 
The Python package 
\texttt{scikit-mobility}~\cite{pappalardo_scikit-mobility_2021}
implements a comprehensive collection of measures suitable for 
human movement data, without combining a subset of them into a single report.
Additionally, the package includes methods for privacy risk assessments; 
however, no privacy mechanisms are 
provided that can be applied to the computation of statistics.

In the context of publishing privacy-preserving aggregated statistics, 
differential
privacy has been proposed for different cases, focusing on
relational data and the interactive setting, \ie, data can only be accessed via 
certain database queries~
\cite{mcsherry_privacy_2009,johnson_towards_2018,dwork_algorithmic_2013}.
For mobility data a multitude of approaches address differential
privacy in the non-interactive setting, \ie, publishing synthetic mobility data
~\cite{chen_differentially_2012,mir_dp-where_2013,fan_differentially_2013,gursoy_utility-aware_2018}.
However, these works assume that a user does not contribute multiple items.
Therefore, we focus on releasing mobility measures to understand the
implications of user-level and item-level privacy.

For mobility data a multitude of approaches address differential
privacy in the non-interactive setting, \ie, the publication of anonymized historical data in their raw format 
by, e.g., generating synthetic data ~\cite{mir_dp-where_2013, bindschaedler_synthesizing_2016, 
gursoy_utility-aware_2018, lestyan_search_2022}. Such data releases are commonly motivated by the 
lack of openly available mobility data for a wide range of (not yet fully specified) use cases. However, often 
the needed analyses are known in advance which allows the application of privacy-preserving techniques 
which are more precise and targeted for the respective use case.

Additionally, many existing approaches assume that a user does not contribute multiple items.
Liu et al.~\cite{liu_learning_2020} studied the utility privacy trade-off for
user-level privacy and compared it to the item-level counterpart.
While they evaluated it in the context of learning discrete distributions, we
focus on a multitude of aggregations comprising a standardized mobility report.
Wilson et al.~\cite{wilson_differentially_2020} present an approach to 
user-level privacy for aggregations by bounding user contribution. They 
account for situations in which the categories used for grouping the data 
are not known a-priori.
However, they guarantee approximate differential privacy.
Aktay et al.~\cite{aktay_google_2020} follow their 
overall approach to compute aggregated mobility
metrics in the context of the COVID-19 pandemic.
They account for user-level differential privacy by bounding user contribution to four records.
We adapt their differential privacy mechanism, 
mainly consisting of adding noise to aggregated statistics 
and bounding user contribution,
and extend it to a variety of mobility measures and evaluate the impact 
of the chosen bound for user contribution.

Amin et al.~\cite{amin_bounding_2019} theoretically analyze the bias-variance
trade-off of bounding user contribution.
Their results show that an optimal contribution limit can be found for which the
expected error of differentially private empirical risk minimization is optimal,
but that there is no contribution bound that is sufficient to eliminate both
bias and variance, even in the limit case of infinite
data~\cite{amin_bounding_2019}.
In this paper, we empirically analyze the bias-variance trade-off of bounding
user contribution.
Additionally, we discuss relevant aspects that determine the trade-off between
item-level and user-level differential privacy and show
consequences of a differentially private mobility report.

\section{Standardized Mobility Report}
\label{sec:standardReport}

Following the idea of a standardized report for general tabular data, 
as implemented by the \texttt{pandas\_profiling} 
package, we
propose a standardized human urban mobility data report that includes 
statistics commonly applied to human mobility data sets~\cite{barbosa_human_2018, 
pappalardo_data-driven_2018, pappalardo_scikit-mobility_2021, luca_deep_2020, cho_friendship_2011}.
This will serve as a basis for practitioners 
to gain fundamental insights from human movement data and establish a basis for further decisions.

\subsection{Mobility data}

Human movement analyses usually focus on stay-points
where a certain amount of time is spent, e.g., \say{home}, \say{work} or
\say{supermarket}~\cite{barbosa_human_2018}, in contrast to way-points that are 
typically only passed by during a trip. In terms of privacy risks, the
stay-points are also of special interest: although there are attacks that use
mobility features such as speed, direction, or 
distance~\cite{rossi_spatio-temporal_2015}, most attacks focus on important
locations~\cite{bennati_privacy_2020}. We will thus 
restrict to a generalized form of human mobility defined as a collection of single trips comprised 
of origin-destination pairs, in particular, we will ignore way-points. This 
results in the following 
requirements for input data 
that are satisfied by many common urban mobility data sets like 
surveys, routing queries, public transport smart card check-ins and -outs, or 
shared mobility rentals.

\begin{figure}[tb]
	\centering
	\input{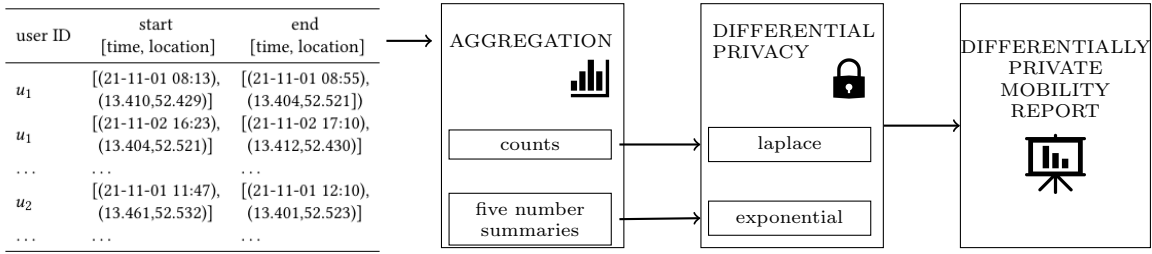}
	\caption{Overview of our process for a differentially private mobility data 
		report.}
	\label{fig:system}
\end{figure}

Let $U$ be a set of user identifiers and $P=\{p=(t,l)\}$ a set of spatio-temporal points $p$
consisting of a timestamp $t$ and a location $l$ referring to 
a given coordinate reference system (CRS), e.g., geographical latitude and longitude. 
A row of the input data set $T$ consists of a user 
identifier $u\in U$ and two spatio-temporal points,
the start and end point, 
that together define a trip $tr$.
We show a mobility data set example in Figure \ref{fig:system}.
Origins and destinations are 
mapped onto a tessellation. This is 
needed to spatially aggregate the data
for statistical analyses, modeling tasks~\cite{luca_deep_2020}, and comprehensive visualizations of trips 
that are not
cluttered with lines~\cite{andrienko_visual_2017}. Typical tessellations are 
administrative boundaries or evenly distributed grids~\cite{luca_deep_2020}. We 
assume that such a 
tessellation is provided for the report.

\subsection{Mobility measures}
Human mobility can mainly be analyzed from three perspectives: places that are
visited, movements between places, and mobility characteristics of individuals
\cite{toch_analyzing_2019}. For \textit{place analyses} the number of movements originating or arriving at 
a location are of interest, while
\textit{trip analyses} focus on where people are coming from or going to. 
The analyses of mobility characteristics of individuals, which we refer to as \textit{user 
analyses}, puts a single person into the center of 
attention. All movements of a person are examined together, for example to evaluate the spatial range 
individuals typically move in. We structure the proposed report in accordance 
with the described analysis perspectives in addition to a general group of \textit{overview analyses}. The 
first three columns of 
Table~\ref{tab:StatisticsOverviewSensitivity} provide an overview of all 
statistics 
included, which are explained in detail below.

The first group of analyses gives an overview of basic information, including the
 number of \textit{trips}, \textit{users} and \textit{locations}.
Temporal properties are examined in three different ways: trip counts 
are aggregated to the 
number of \textit{trips over time}, where the time interval is chosen 
automatically to be a day, week or month 
depending on the time range covered by the data. 
This measure reveals the temporal coverage and completeness of the data, 
the potential existence of (un)intentional interruptions
and patterns of seasonality. 
It is provided as trip counts per time interval and as a five-number summary.
Weekly and daily cycles, which capture patterns that are 
characteristic for human mobility \cite{schlich_habitual_2003, pappalardo_data-driven_2018}, can be 
inferred from the 
number of \textit{trips per weekday} and \textit{trips per hour} disaggregated by weekday and weekend, 
respectively.

The second group, place analysis, provides 
information on the geospatial distribution of the data in terms of the 
provided tessellation. The main measure computes 
the total number of \textit{visits per location}
over the entire time period of observation. 
As points of trips can potentially lie outside the tessellation, the number of 
outliers is provided as well.
Note that there are twice as many visits as trips, since both start 
and end point are considered. 

The frequency of location visits usually differs depending on the time of day 
\cite{pappalardo_data-driven_2018}. Therefore, spatial
distributions are also disaggregated as \textit{visits per destination by time} and split by weekday and 
weekend. 
To limit the number of outputs, we use a default time span of four hours to create six time windows 
unlike, e.g., hourly time windows that would result in 24 views. The default time windows are defined as 
follows: 
2:00 - 6:00, 6:00 - 10:00, 10:00 - 14:00, 14:00 - 18:00, 18:00 - 22:00, 22:00 - 02:00.
Only destinations are considered to correctly represent spatio-temporal patterns. 
For illustration, consider the following example: in the morning
people commute from the outskirts to the city center and vice
versa in the evening. If both origins and destinations would be considered, the
outskirts and the city center would be highly frequented in the morning and
 evening, thus not revealing important temporal patterns.

The third group covers trip analyses. Trips are aggregated and counted according to 
origin and destination tiles, resulting in an OD matrix, named \textit{OD flows}.
Additionally, the \textit{travel time} and \textit{jump 
length}~\cite{luca_deep_2020} 
are computed, the latter representing the geographical straight-line distance between the trip's origin and 
destination. They are reported in form of five-number summaries and histograms, 
i.e., counts. 
Travel time and jump length typically decay as a 
power law~\cite{brockmann_scaling_2006-1} with many short times and distances 
and few very long trips. 
An unbound histogram would therefore be highly right-skewed, either with many small bins or a few 
non-expressive bins at the lower end of the histogram. To limit the number of histogram bins for a more 
comprehensive representation, we propose to cut them off at a defined maximum based on domain 
knowledge and to additionally provide the number of outliers above the defined threshold. 
In addition to usability reasons, a pre-defined maximum helps to control the privacy-utility trade-off as 
will be shown in Section~\ref{sec:dpReport}. 

The last group on user analyses compiles the following measures, 
summarized as histograms and five-number summaries. The number of \textit{trips 
per user} 
computes a user's overall contribution to the data set.
\textit{Locations per user} represents the distinct locations visited by a user 
\cite{pappalardo_data-driven_2018} 
and describes the diversity of locations a user visits. The \textit{mobility entropy} entails related 
information: it is defined as the Shannon entropy of the user's visits which quantifies the probability of 
predicting a user's whereabouts~\cite{song_limits_2010, 
pappalardo_data-driven_2018, luca_deep_2020}. 
The \textit{radius of gyration} is the characteristic distance traveled by a user 
\cite{gonzalez_understanding_2008, 
pappalardo_data-driven_2018, luca_deep_2020}, computing the spread of all locations visited 
by an individual around their center of mass. Again, the distribution of radii of gyration 
follows a power law, thus setting cut off values for the histogram is beneficial for usability and privacy 
reasons. 
Further entailed is the \textit{time between trips}  which
computes the amount of time between the end time of a user's trip and
 the beginning of the next one. It thus quantifies the temporal density of user trips in the data, 
i.e., if user movements are recorded every other hour, day, week or month. 

\begin{table}[tb]
	\small\sf\centering
	\caption{Overview of mobility measures and sensitivity.}
	\label{tab:StatisticsOverviewSensitivity}
	\begin{tabularx}{\textwidth}{lp{3.4cm}lX}
		\toprule
		Group & Measure & Function & Sensitivity \\
		\midrule
		\multirow{6}{*}{Overview}
		& trips& $\mathrm{count}_{tr}$ & $M$ \\
		& users&$\mathrm{count}_{u}$&1 \\
		& locations&$\mathrm{count}_{p}$&$2\cdot M$ \\
		& trips over time&$\mathrm{counts}_{tr}$,five-number 
			summary$_{tr}$&$M$, $M$ \\
		& trips per weekday&$\mathrm{counts}_{tr}$&$M$ \\
		& trips per hour & $\mathrm{counts}_{tr}$&$M$ \\
		\midrule
		\multirow{2}{*}{Place analysis}
		& visits per location&$\mathrm{counts}_{p}$ &$2\cdot M$\\
		& visits per destination and time&
		$\mathrm{counts}_{tr}$&$M$ \\
		\midrule
		\multirow{3}{*}{Trip analysis}
		& OD flows&$\mathrm{counts}_{tr}$&$M$\\
		& travel time&$\mathrm{counts}_{tr}$, five-number 
			summary$_{tr}$&$M$, $M$ \\
		& jump length&$\mathrm{counts}_{tr}$, five-number 
			summary$_{tr}$&$M$, $M$ \\
		\midrule
		\multirow{6}{*}{User analysis}
		& trips per user&$\mathrm{counts}_{u}$, five-number 
			summary$_u$&$1, 1$ \\
		& radius of gyration&$\mathrm{counts}_{u}$, five-number 
			summary$_u$&$1, 1$\\
		& locations per user&$\mathrm{counts}_{u}$, five-number 
			summary$_u$&$1, 1$\\
		& mobility entropy&$\mathrm{counts}_{u}$, five-number 
			summary$_u$&$1, 1$\\
		& time between trips& 
		$\mathrm{counts}_{tr}$, five-number summary$_{tr}$&$M$,$M$ \\
		\bottomrule
	\end{tabularx}
\vspace*{-1em}
\end{table}

\section{Differentially Private Mobility Report}
\label{sec:dpReport}

\subsection{System Model and Design}

Before describing our system for a differentially private mobility data report,
we introduce respective definitions and notations used throughout this paper and
illustrate the intermediate steps to such a report as depicted in 
Figure~\ref{fig:system}.
Assume, a company owns a mobility data set $T$ and needs to release a
mobility report.
For a standardized mobility data report, statistics
defined in Section~\ref{sec:standardReport} are computed and visualized.
Despite the aggregation of data, the mobility report consists of summary
statistics which are vulnerable to
reconstruction attacks~\cite{dwork_algorithmic_2013}.
By observing answers from measures, such as listed in 
Table~\ref{tab:StatisticsOverviewSensitivity}, an attacker can recover secret
information, such as trips.

Differential privacy provides mathematical guarantees for the privacy of an
individual~\cite{dwork_calibrating_2006}.
To define differential privacy we need the notion of neighboring data sets.
\begin{definition}[Neighboring data sets]
	\textit{Two mobility data sets $T_1$ and $T_2$ are neighbors
		if they differ in exactly one row, \ie one
		data set is obtained from the other by removing or adding one row.}
\end{definition}

\begin{definition}[Differential Privacy]
	\textit{An algorithm~$\mathcal{A}$ provides $\varepsilon$-differential 
		privacy if all neighboring data sets~$T_1$ and $T_2$ and all
		outputs~$O\subseteq \text{Range}(\mathcal{A})$ satisfy}
	$P[\mathcal{A}(T_1) \in O] \,\leq\, \mathrm{e}^{\varepsilon} \cdot 
	P[\mathcal{A}(T_2) \in O]$.
\end{definition}
We denote a randomized algorithm by $\mathcal{A}$
that takes a mobility data set $T$ as input and
outputs a value from some output space $\textit{Range}(\mathcal{A})$.
The concept of differential privacy is that the output of an
algorithm~$\mathcal{A}$ remains
nearly unchanged if the data of one individual is removed or added.
In this way, differential privacy limits the impact of a single individual on
the analysis outcome, preventing the reconstruction of an individual's data.
The parameter $\varepsilon$ captures the privacy loss and determines
how similar the randomized outputs are based on $T_1$ and $T_2$, and thus
determines the impact of a single individual's data point to the output.

\subsection{User-Level Privacy}
\label{sec:userlevel}
Usually, in the definition of differential privacy, neighboring data sets 
differ in
one row, \ie in one item. Thus, it is referred to as item-level 
privacy~\cite{dwork_calibrating_2006}.
However, in practice and especially with mobility data, every user may contribute
multiple items or trips. Therefore, the definition of neighboring data
sets needs to be adjusted with respect to user-level privacy.
In user-level privacy, two mobility data sets $T_1$ and $T_2$ are neighbors
if they differ in the data of a single user.
If users have only one item, item-level is equal to user-level privacy.

An important notion when dealing with differential privacy and differentiating
between item-level and user-level privacy is the sensitivity of a
function~\cite{dwork_calibrating_2006}.
The sensitivity of a function~$f$ is the maximum difference that an output can
change by removing or adding a row (item-level) or a
user (user-level).
\begin{definition}[Sensitivity]
	\label{sense}
	\textit{The $L_1$-sensitivity of $f$ is defined as}
	$\Delta f = \max_{(T_1,T_2)} \vert f(T_1) - f(T_2)\vert_{1}$
	\textit{for any two neighboring mobility data sets $T_1$ and $T_2$.}
\end{definition}

In our mobility report, most of the functions, listed in
Table~\ref{tab:StatisticsOverviewSensitivity}, are counts
and as such output numeric values.
For example, the function \textit{count} outputs a single number, while
\textit{counts}
outputs a number for each category and bin, respectively.
A common mechanism for numeric functions is the Laplace
mechanism, where calibrated noise is added to the function's output,
drawn from a Laplace distribution~$Lap()$~\cite{dwork_calibrating_2006}.
\begin{definition}[Laplace mechanism]
	\label{laplace}
	\textit{Let $f:T^n\to \mathbb{R}^k$ with arbitrary domain $T$.
		The Laplace mechanism is defined as}
	$\mathcal{A}(T) = f(T)+ (Y_1,\ldots,Y_k)$
	where $Y_i$ is a random variable drawn from $ Lap\left(\Delta f / 
	\varepsilon\right) $ with mean $0$ and variance $2(\Delta
	f / \varepsilon)^2$.
\end{definition}
Note that for \textit{counts}
$k$ equals the
number of categories/bins, \eg, locations, while $k=1$ 
for the \textit{count} function.
The magnitude of noise is calibrated according to the sensitivity $\Delta f$ of
a function.

If we assume item-level privacy for our report, this means for the number of
\textit{trips over time} that noise with $\Delta f=1$ is added to each count,
since removing or adding one trip changes the count by one.
While this effectively hides the presence of one trip of an individual,
it does not protect the presence/absence of an individual with multiple trips.
In other words, removing or adding all trips of a user changes the number
of \textit{trips over time} by the amount of trips of the corresponding user
instead of just one.

Suppose we are interested in the number of \textit{OD flows}.
A user has made 10 trips. If we remove the user, the counts will always change by 10: 
If the user made all the trips between the same origin and destination, the number for this
OD flow changes by 10. The other counts remain unchanged.
If the user made the 10 trips between varying OD pairs, the counts for each respective of pair
change by 1. This means that the maximum influence of a user on such a 
measure $f$ and therefore its sensitivity $\Delta f=10$.
In practice, however, a user can contribute an arbitrary number of trips and
the sensitivity of $f$ is thus unbounded.

\subsection{Bounded user contribution}
\label{sec:boundedContr}
In this section, we present the differentially private functions of our
mobility measures proposed in Section~\ref{sec:standardReport}.
We list the functions and sensitivity to guarantee user-level privacy in
Table~\ref{tab:StatisticsOverviewSensitivity}.

In order to limit the sensitivity,
we need to limit the number of possible trips $M$ of a user.
If we choose the highest number of trips a user has in our data set for $M$, we
assume local sensitivity.
Local sensitivity depends on the data set and not only on the function.
Therefore, the local sensitivity or the maximum number of trips may jeopardize
the privacy of a user.

Through sampling, the number of trips of a single user are bounded according to
$M$ and remaining records are removed.
The places people visit is highly predictable for most individuals
~\cite{gonzalez_understanding_2008, song_limits_2010}.
There are only few places that a person visits regularly,
usually the home and work place, which make up the majority of locations in a
person's mobility pattern~\cite{do_places_2014}.
Therefore, we assume that the global geospatial patterns of a mobility data set
remain intact even though only a small sample of each person's trips are
included.

The question arises as to what maximum the number of trips should optimally be 
set at. For example, Aktay et al.~\cite{aktay_google_2020} have limited the 
number to
4 in their differentially private mobility report, without explaining this
further. We will look at this question in detail as part of our evaluation.

\subsubsection{Counts}
The count functions based on users ($\textit{count}_{u}$) have a
sensitivity of $1$, since removing a user, no matter how many trips they make,
only changes one count by $1$. E.g., the \textit{trips per user} are
represented with a histogram, each bin representing a possible number of trips.
Removing/adding a user with their trips will only change the count for one bin
by $1$.
The count functions based on trips ($\text{count}_{tr}$) have a sensitivity of
$M$ as explained in the previous example.
Since a trip consists of two points (start and end),
the count functions based on points ($\text{count}_{p}$) have a sensitivity
of~$2\cdot M$.

To this end, we guarantee differentially private counts. However, \eg,
visited locations or reported categories can reveal the identity of an 
individual.
For example, if only tiles that were actually visited are included in the
report, we can infer that all reported tiles were visited at least
once, while all tiles that were not included are not part of the data set.
Geo-indistinguishability~\cite{andres_geo-indistinguishability_2013}
has been proposed to protect exact locations.
In geo-indistinguishability two locations within a given radius are
indistinguishable.
However, based on the specified radius, this definition reveals some
information.
For example, the fact that an individual has visited a particular district may
be known, even if the exact location is randomized.
Therefore, the privacy within a radius is crucial.
We assume that the fact that users have visited cells in the
given tessellation does not jeopardize privacy.
Consequently, we consider the tessellation as given to obtain a finite
countable number of geometric shapes.
In our proposed mobility report, noisy counts are only displayed
for the locations included in the tessellation while all other points are
summarized as a single noisy outliers count.

The tessellation defines categories for the spatial dimension.
Categories for temporal dimensions, which is reflected by 
\textit{trips over time}, are aggregated to specific time intervals, such as
day, week or month.
Empty intervals in between should be filled with 0 values so that noise can be
applied as well. 
While the tessellation also provides a fixed bounding of the spatial extent,
there is another issue for \textit{trips over time}:
revealing the exact first and last day of the entire time interval violates the
privacy. The same issue of leaking minimum and maximum values applies to any
count-based measure that has no 
pre-defined categories or bins, namely, 
\textit{travel time}, \textit{jump length}, \textit{trips per user}, \textit{time between trips}, \textit{radius of 
gyration}, \textit{locations per user} and \textit{mobility entropy}. 
Similar to the given tessellation we can cut off bins at a defined 
minimum and maximum based on domain knowledge.
Data points outside this interval are summarized as a single noisy outliers
count.
Instead, we can also determine the minimum and maximum differentially private
to obtain bounds.
Since the minimum and maximum are included in the five-number summary we do not
need to compute them twice.

\subsubsection{Five-Number Summary}
The five-number summary can be returned differentially private with
the exponential mechanism~\cite{mcsherry_mechanism_2007}.
This mechanism does not add noise to
the output. Instead it returns the best answer from a set based on a scoring function.
Differential privacy is 
guaranteed since sometimes an output is returned even though it has not the highest 
score. 
\begin{definition}[Exponential mechanism]
	\label{exponential}
	\textit{Given an input data set $T$ the Exponential 
	mechanism~\cite{mcsherry_mechanism_2007}
		randomly samples
		an output $O\subseteq \text{Range}(\mathcal{A})$ with a
		probability	proportional to}
	$e^{\frac{\varepsilon s(T,O)}{2\Delta s}}$,
	where $s$ is the scoring function and $\Delta s$ the corresponding
	sensitivity.
\end{definition}

We use the exponential mechanism to determine the five-number summary including
the minimum and maximum. In this case the scoring function is a rank function of
the sorted input.
Since we determine the index of an element, a user with $M$ trips influences
the output by $M$. Therefore, the sensitivity for the five-number summary in
Table~\ref{tab:StatisticsOverviewSensitivity} is the same as that for 
counts.
Note that the element returned by the exponential mechanism is always a member
of the set $\text{Range}(\mathcal{A})$.
This is reasonable for a finite set where a noisy response is not useful.
Therefore, a pre-defined minimum/maximum is more privacy-preserving
and should be preferred as far as possible.

\vspace{1cm}
\section{Application and Evaluation}
\label{sec:eval}

\subsection{Implementation and Provision of Code}
The differentially private mobility data 
report proposed in Section \ref{sec:dpReport} is implemented as a 
python code package\footnote{Link to package repository: 
\url{https://github.com/FreeMoveProject/dp_mobility_report}}
and provided as 
open source code under the MIT license. The implementation is inspired by \newline the 
\texttt{pandas\_profiling} package. It expects a \textit{pandas 
DataFrame}~\cite{mckinney-proc-scipy-2010} in the required format 
as input and outputs a report as an 
HTML file. This 
serves as an initial step to provide an easy to use implementation of the report in practice. We 
envision to further develop the package to comply with common standards and to adjust functionality 
based on testings with practitioners.

\subsection{Experimental Setup}
In the following, we compare the deviation of the mobility report from differentially private variants. For this 
purpose, we focus on  
four selected mobility measures, introduced in 
Section~\ref{sec:standardReport}, that can 
be considered typical and unique for mobility data: \textit{number of trips}, \textit{visits per location}, 
\textit{OD flows} and \textit{radius 
of gyration}. For a full analysis of error 
measures for the entire report we refer the reader to our online repository\footnote{Link to 
evaluation repository: 
\url{https://github.com/FreeMoveProject/evaluation_dp_mobility_report}}.

Below, we describe the data sets, the choice of parameters for the privacy guarantees, and the error 
measures to quantify the deviations.

\subsubsection{Data sets}
We use three data sets, denoted by {\tt 
MADRID}, {\tt GEOLIFE} and {\tt BERLIN}, commonly used in research and applications on human mobility. 
The first two are real, open data sets; {\tt BERLIN} is a synthetic data set created with a traffic simulation 
software.

The data set
{\tt MADRID}~\cite{crtm_encuesta_2018, 
consorcio__regional__de__transportes_de_madrid_documento_2019} is a 
survey
on the mobility behavior of 75,208 residents of the Community of Madrid.
Participants were asked about each trip they made on one weekday (Monday to Thursday) they selected 
between February and May 2018.
The data set contains a tessellation consisting of 1,259 irregular 
sized traffic cells~\cite{crtm_zonificacion_2018} that serve as the possible 
origins and destinations of the 
survey. 2,007 points were 
outside the given tessellation.
As our input specification requires coordinates for origins 
and destinations, we replace the tile IDs with the coordinates of the respective tile centroids.

{\tt GEOLIFE} was collected and released as part of 
the eponymous project 
\cite{zheng_understanding_2008, zheng_mining_2009, zheng_geolife_2010} in which movements of 182 
participants were 
recorded between 2007 and 2012 using GPS devices. Users were not continuously tracked over 
the entire time period; instead, single trips were recorded when users actively started tracking. 
Therefore, the time period and number of recorded trips differ greatly between individual users.
Nine users contribute almost half of the trips with more 
than 400 trips each, while half of all users recorded only 27 trips or less. As we only use start and 
end locations, all way-points of the trips are removed for our purposes. 
Most traces are located in Beijing, China. As no tessellation is provided, we use a hexagonal grid 
based on the H3 geospatial indexing system\footnote{\url{https://eng.uber.com/h3/}} that covers the 
majority of 
data points within the Beijing center 
with an H3 resolution of 7, omitting 3,583 (9.5\,\,\%) of the records.

The {\tt BERLIN} data set is
produced by open-source traffic simulation software 
TAPAS~\cite{heinrichs_introduction_2017} of
German 
Aerospace Center
and has been calibrated to the mobility behavior of the population of Berlin on 
a 
typical workday based
on a survey in 2017. We work with a 10\,\% sample of the entire Berlin 
population which results in 378,759 
users and 1,35 M trips and use traffic cells
of Berlin~\cite{berlin_verkehrszellenteilverkehrszellen_2014} as a 
tessellation. 

\begin{table}[tb]
	\small\sf\centering
	\caption{Overview of evaluation data sets}
	\label{tab:DataSets}
	\begin{tabular}{lrrrrrrrrr}
		\toprule
		&&& \multicolumn{3}{c}{Tessellation} & \multicolumn{4}{c}{Trips per 
			user} \\
		\cmidrule(l){4-6}
		\cmidrule(l){7-10}
		&Trip count&User count&Tile count&Tessellation area&Mean tile 
		area&Max&Median&Mean&SD\\
		\midrule
		{\tt GEOLIFE} & 18,670 & 182 &  962 & 3,566  $km^2$  & 3.7 $km^2$ & 
		2,153 & 27.5 & 102.6 & 250.0 \\
		{\tt MADRID} & 222,744 & 75,208 &  1,259 & 8,031 $km^2$ & 6.4  $km^2$ & 
		20 & 2.0 & 3.0 & 1.5 \\
		{\tt BERLIN}  &1,417,134 & 378,759 &  386 & 891 $km^2$ & 2.3 $km^2$ & 
		16 & 4.0 
		& 3.7 & 1.8\\
		\bottomrule
	\end{tabular}
\end{table}

\begin{figure}[tb]
	\begin{subfigure}[t]{.3\textwidth}
		\centering
		\includegraphics[width=0.9\textwidth]{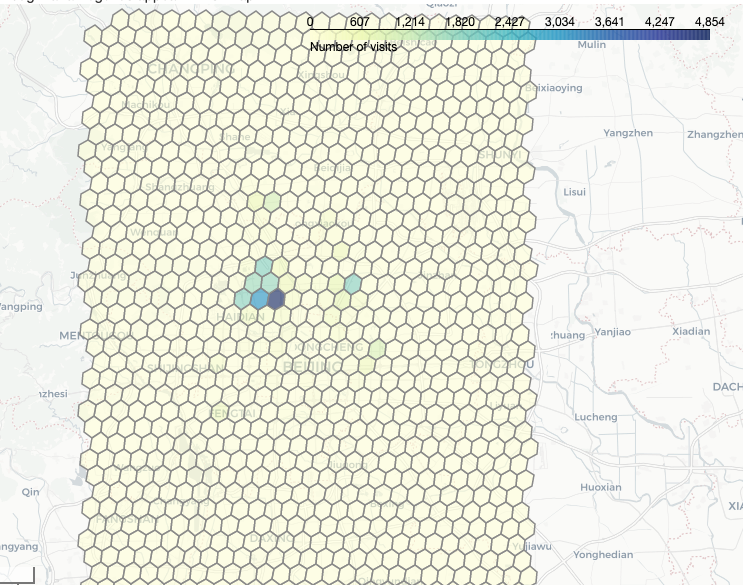}
		\caption{\tt GEOLIFE}
	\end{subfigure}%
	\begin{subfigure}[t]{.3\textwidth}
		\centering
		\includegraphics[width=0.9\textwidth]{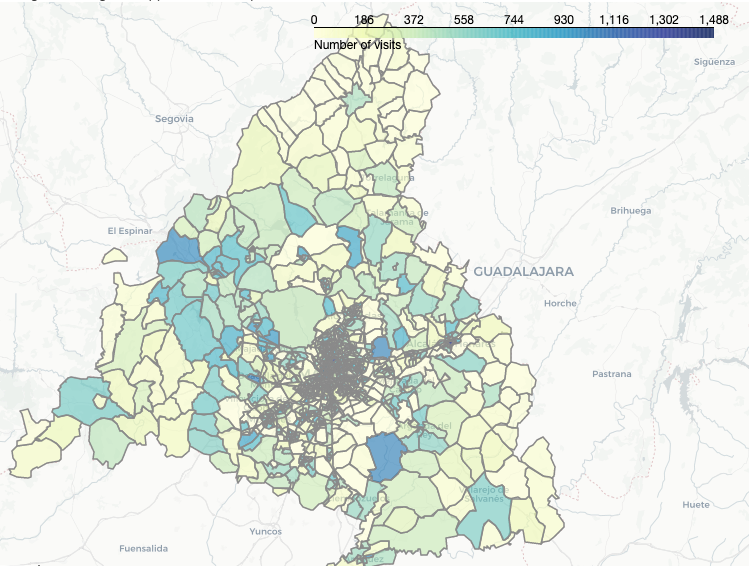}
		\caption{{\tt MADRID} (Tessellation by
			~\cite{crtm_zonificacion_2018})}
	\end{subfigure}
	\begin{subfigure}[t]{.3\textwidth}
		\centering
		\includegraphics[width=0.9\textwidth]{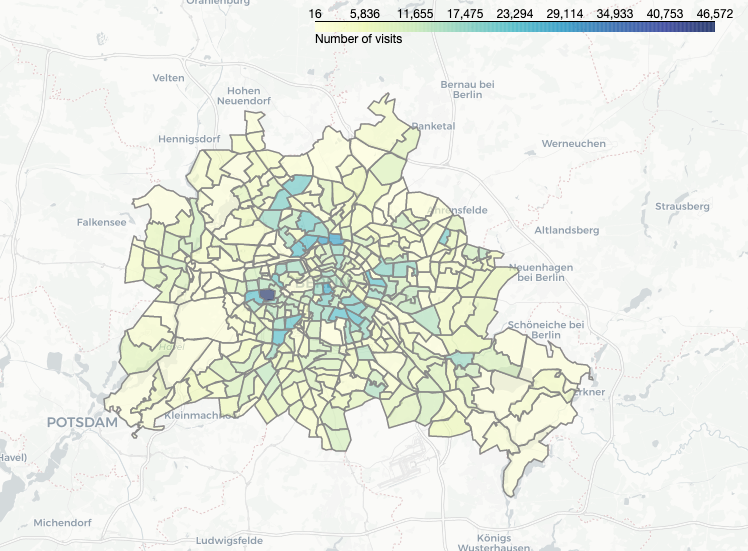}
		\caption{{\tt BERLIN} (Tessellation by
			~\cite{berlin_verkehrszellenteilverkehrszellen_2014})}
	\end{subfigure}
	\caption{Tessellation and visits per location }
	\centering\small\textsuperscript{Basemap: \textcopyright OpenStreetMap contributors 
		\textcopyright CartoDB}
	\label{fig:map}
\end{figure}

In Table \ref{tab:DataSets} we show the main characteristics of the selected
data sets and in Figure~\ref{fig:map} we
visualize their tessellations and the mobility measure of \textit{visits per 
location}.
Particularly relevant for consideration of user contribution bounds is the difference in user contributions 
per data set: while the distribution of trips per user in {\tt GEOLIFE} is strongly skewed to the right, 
those of {\tt MADRID} and even more {\tt BERLIN} are rather balanced with a larger number of users 
and trips and a lower contribution per user. {\tt BERLIN} has by far the most users and trips distributed on 
the smallest geospatial area.

\subsubsection{Evaluation Runs}
To evaluate the similarity of mobility measures with and without privacy guarantees we use different 
values of the maximal user contribution $M$ and the privacy budget 
$\varepsilon$.
The choices of $M$ are based on the distribution 
of the number of trips $M_u$ per user $u$ for a given data set: quartiles, the 
10th percentile,
the 90th percentile, the maximum and the minimum ($=$ single trip per user), 
resulting in seven variants if 
no values overlap (e.g., the 10th percentile, second and third quartile have a 
value of 2 for {\tt 
MADRID}). For a given $M$, a random sample of $\min (M, M_u)$ records is drawn per $u$.

For comparison, we also include runs that only apply sampling and do not add any noise, which we 
refer to as \textit{withoutDp}.
For simplicity, the privacy budget is not split between the mobility measures, as it should be for a report 
granting overall $\varepsilon$-differential privacy. 
Instead we use the same budget for each measure which would sum up 
for an entire report.
For evaluation purposes, all combinations of $M$ and $\varepsilon$ are run 10
times.
In order to present possible deviations, we add error bars to show the variance.

\subsubsection{Error measures}
We use the following measures to quantify the resemblance between mobility data reports without 
and with differential privacy guarantees. For 
any mobility measure $x$ the differentially private counterpart is denoted as 
$x'$.
The higher the resemblance the lower the error which suggests a better utility 
of the mobility data report.

\textbf{TripCountError.}
We use the relative error to quantify the deviation of total trip counts $\mathrm{count}_{tr}$:

$$\mathrm{TripCountError} := \frac{|\mathrm{count}_{tr} - 
\mathrm{count}_{tr}'|}{\mathrm{count}_{tr}}\text{.}$$

\textbf{LocationError.}
The Earth Mover's Distance (EMD)~\cite{rubner_earth_2000} is used to evaluate 
the deviation of visit 
counts in each tile. The EMD determines the least amount of work necessary to reshape one 
distribution to another. Let $L$ be the distribution of visit counts per tile. The EMD between 
$L$ and $L'$, denoted as 
$\mathrm{LocationError}$, is defined as:

$$\mathrm{LocationError} := \mathrm{EMD}(L, L') = \frac{\sum_{i = 1}^n \sum_{j = 1}^n d_{ij} f_{ij}}{\sum_{i 
= 1}^n \sum_{j = 1}^n 
f_{ij}},$$

where $n$ is the number of tiles, $d_{ij}$ is the distance between the $i$-th and $j$-th tile and $f_{ij}$ 
denotes proportion of flows between the two tiles 
required to transform $L$ to $L'$.
We use the haversine distance between the centroids of tile $i$ and $j$ to determine the distance 
$d_{ij}$, allowing us to quantify geospatial shifts, where shifts between neighboring tiles are 
weighted less than shifts between distant tiles. Thus, we can intuitively interpret the resulting 
$\mathrm{LocationError}$: A value of, e.g., 100 means that every point in $L$ needs to move 100 meters 
on 
average to reproduce $L'$. Note that we thereby do not evaluate changes in absolute visit counts in 
tiles, but changes of relative shares.

\textbf{ODFlowError.}
We use the symmetric mean absolute percentage error (SMAPE) to measure the divergence between 
matrices representing origin-destination flows. First, the \textit{OD flows} are 
normalized by dividing 
through the sum of \textit{OD flows}, 
so it is not accounted for changes in the absolute count through sampling but only for changes in the 
distribution. The normalized matrix is denoted as $A$. 
To ensure that the error measure is independent of the size of the tessellation, only the combined 
support of $A$ and $A'$ is considered. Thus, $n$ is the number of combinations 
where at least one 
matrix cell of $A$ or $A'$ is not 0. Otherwise, the error could always be reduced by increasing the 
tessellation and thereby the percentage of 0 valued cells. 
The absolute percentage is determined and averaged to a single measure 
defined as 

$$\mathrm{ODFlowError}:=\frac{2}{n} \sum_{a_c + a'_c>0} 
\frac{|a_c - a'_c|}{(a_c + a'_c)},$$

where $a_c$ denotes the entry of the OD matrix $A$ at index $c$. Note that 
$\mathrm{ODFlowError}$ has a lower bound of 0 and an upper bound of 2.

\textbf{RadiusOfGyrationError.}
The error for the radii of gyration makes use of quartiles and is
computed using SMAPE. 
The distribution of the radii of gyration, consisting of the five-number 
summary, 
is denoted by $\mathrm{ROG}$:

$$\mathrm{RadiusOfGyrationError} := 
\frac{2}{5} \sum_{q= 1}^5 
\frac{|\mathrm{ROG}_q - \mathrm{ROG}'_q|}{(\mathrm{ROG}_q + 
\mathrm{ROG}'_q)}\text{,}$$
where $q$ is the index of the five-number summary value in $\mathrm{ROG}$.

\subsection{Item level vs. user level privacy}
\label{sec:item_vs_user_level}
There is a substantial difference between
item-level privacy and user-level privacy for mobility data, as they typically
contain multiple items per user. In the following, we denote the maximum
number of trips a user contributes to the data set by $M$, and thus the
sensitivity to user-level privacy.
Note that in real applications, the provision of $M$
should also be differentially private, unless it is externally set and enforced
by sampling techniques~\cite{amin_bounding_2019}.

\begin{table}[tb]
	\small\sf\centering
	\caption{Error measures for user-level vs. item-level privacy with 
	$\varepsilon = 1$. $M$ equals 
		the maximum user contribution per data set ({\tt GEOLIFE}: $2,153$; 
		{\tt MADRID}: $20$; {\tt BERLIN}: $16$.) }
	\label{tab:UserVsItem}
	\vspace*{-1em}
	\begin{tabular}{lrrrrrr}
		\toprule
		& \multicolumn{2}{c}{{\tt GEOLIFE}} & \multicolumn{2}{c}{{\tt MADRID}} &
		\multicolumn{2}{c}{{\tt BERLIN}} \\
		\cmidrule(l){2-3}
		\cmidrule(l){4-5}
		\cmidrule(l){6-7}
		error measure & user-level & item-level & user-level & item-level & 
		user-level & 
		item-level \\
		\midrule
		TripCountError			& 0.14 & 0.00 & 0.00 & 0.00 & 0.00 & 0.00 \\
		LocationError 			& 17,142.48 & 353.50 & 202.14 & 18.87 & 10.32 & 
		1.23 \\
		OdFlowError 			& 2.00 & 1.99 & 1.96 & 1.78 & 1.44 & 0.60 \\
		RadiusOfGyrationError 	& 0.50 & 0.35 & 0.00 & 0.00 & 0.08 & 0.08 \\
		\bottomrule
	\end{tabular}
\end{table}

We provide the errors for item-level and user-level
privacy for $\varepsilon = 1$ in Table~\ref{tab:UserVsItem}. It is hardly
surprising that item-level privacy, which provides weaker privacy guarantees,
results in less dissimilarity to the original data set than user-level privacy.
However, the difference between the privacy levels depends heavily on $\varepsilon$, the
data set, and the mobility measure.

The $\mathrm{LocationError}$
illustrates well that 
user-level privacy causes an error several times higher than item-level
privacy, in particular for {\tt GEOLIFE}
(48 times for {\tt GEOLIFE} vs. 11 times for {\tt MADRID}),
which is likely due to the substantially larger sensitivity. 
On the other hand, adding noise to the total number of trips has almost no impact
for either privacy setting. This is to be expected since the measure consists of
a single, typically high absolute number that is more robust to noise. Again,
for {\tt GEOLIFE}, the high sensitivity strongly affects user privacy guarantees and
results in a relative error of 14\,\% even for this top-level aggregation.
On the other hand, there is no relevant difference w.r.t. the 
$\mathrm{RadiusOfGyrationError}$, as its sensitivity equals 1 and 
independent of $M$ (the latter 
holds for most other user analyses,
c.f. Table \ref{tab:StatisticsOverviewSensitivity}).
The difference for {\tt GEOLIFE} is likely random, as a standard deviation of 
$\sigma=0.20$ for item-level and $\sigma=0.14$ for user-level indicates.

Finally note that, with the exception of {\tt BERLIN}, even item-level
privacy yields an error of more than 100\,\% for \textit{OD flows}, suggesting 
that such
analyses require $\varepsilon>1$ to produce useful results. This
is due to the fact that the number of origin-destination combinations increases quadratically
with the number of tiles, while the corresponding counts are much
smaller and thus more sensitive to noise.

\subsection{Interplay between the upper bound of trips per user $M$ and privacy guarantee 
$\varepsilon$}
\label{sec:interplay}

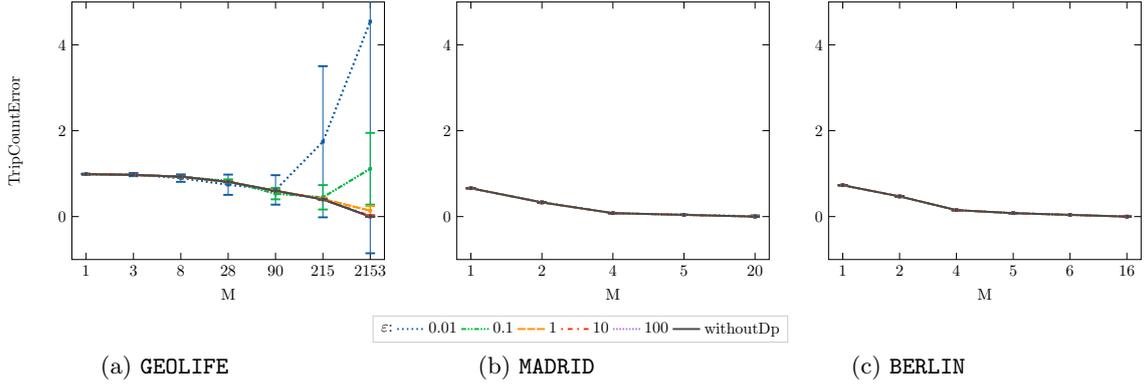
\begin{figure}[tb]
	\centering
	\begin{minipage}[t]{0.3\textwidth}	
\begin{tikzpicture}[scale=0.6]

\definecolor{color0}{rgb}{0.0470588235294118,0.364705882352941,0.647058823529412}
\definecolor{color1}{rgb}{0,0.725490196078431,0.270588235294118}
\definecolor{color2}{rgb}{1,0.584313725490196,0}
\definecolor{color3}{rgb}{1,0.172549019607843,0}
\definecolor{color4}{rgb}{0.717647058823529,0.454901960784314,0.83921568627451}

\begin{axis}[
legend cell align={left},
legend columns=6,
legend style={
  fill opacity=0.8,
  draw opacity=1,
  text opacity=1,
  at={(1.1,-0.15)},
  anchor=south east,
  draw=white!80!black
},
tick pos=both,
x grid style={white!69.0196078431373!black},
xlabel={M},
xmin=-0.3, xmax=6.3,
xtick style={color=black},
xtick={0,1,2,3,4,5,6},
xtick={0,1,2,3,4,5,6},
xtick={0,1,2,3,4,5,6},
xtick={0,1,2,3,4,5,6},
xtick={0,1,2,3,4,5,6},
xtick={0,1,2,3,4,5,6},
xtick={0,1,2,3,4,5,6},
xtick={0,1,2,3,4,5,6},
xtick={0,1,2,3,4,5,6},
xtick={0,1,2,3,4,5,6},
xtick={0,1,2,3,4,5,6},
xtick={0,1,2,3,4,5,6},
xtick={0,1,2,3,4,5,6},
xtick={0,1,2,3,4,5,6},
xtick={0,1,2,3,4,5,6},
xtick={0,1,2,3,4,5,6},
xtick={0,1,2,3,4,5,6},
xtick={0,1,2,3,4,5,6},
xtick={0,1,2,3,4,5,6},
xtick={0,1,2,3,4,5,6},
xtick={0,1,2,3,4,5,6},
xtick={0,1,2,3,4,5,6},
xtick={0,1,2,3,4,5,6},
xtick={0,1,2,3,4,5,6},
xticklabels={1,3,8,28,90,215,2153},
xticklabels={1,3,8,28,90,215,2153},
xticklabels={1,3,8,28,90,215,2153},
xticklabels={1,3,8,28,90,215,2153},
xticklabels={1,3,8,28,90,215,2153},
xticklabels={1,3,8,28,90,215,2153},
xticklabels={1,3,8,28,90,215,2153},
xticklabels={1,3,8,28,90,215,2153},
xticklabels={1,3,8,28,90,215,2153},
xticklabels={1,3,8,28,90,215,2153},
xticklabels={1,3,8,28,90,215,2153},
xticklabels={1,3,8,28,90,215,2153},
xticklabels={1,3,8,28,90,215,2153},
xticklabels={1,3,8,28,90,215,2153},
xticklabels={1,3,8,28,90,215,2153},
xticklabels={1,3,8,28,90,215,2153},
xticklabels={1,3,8,28,90,215,2153},
xticklabels={1,3,8,28,90,215,2153},
xticklabels={1,3,8,28,90,215,2153},
xticklabels={1,3,8,28,90,215,2153},
xticklabels={1,3,8,28,90,215,2153},
xticklabels={1,3,8,28,90,215,2153},
xticklabels={1,3,8,28,90,215,2153},
xticklabels={1,3,8,28,90,215,2153},
y grid style={white!69.0196078431373!black},
ylabel={TripCountError},
ymin=-1, ymax=5,
ytick style={color=black}
]
\path [draw=color0, line width=0.04pt]
(axis cs:0,0.972)
--(axis cs:0,1);

\path [draw=color0, line width=0.04pt]
(axis cs:1,0.941)
--(axis cs:1,1.015);

\path [draw=color0, line width=0.04pt]
(axis cs:2,0.809)
--(axis cs:2,0.983);

\path [draw=color0, line width=0.04pt]
(axis cs:3,0.506)
--(axis cs:3,0.978);

\path [draw=color0, line width=0.04pt]
(axis cs:4,0.278)
--(axis cs:4,0.964);

\path [draw=color0, line width=0.04pt]
(axis cs:5,-0.0169999999999999)
--(axis cs:5,3.503);

\path [draw=color0, line width=0.04pt]
(axis cs:6,-0.854)
--(axis cs:6,9.94);

\addplot [very thick, color0, mark=-, mark size=3, mark options={solid}, only marks, forget plot]
table {%
0 0.972
1 0.941
2 0.809
3 0.506
4 0.278
5 -0.0169999999999999
6 -0.854
};
\addplot [very thick, color0, mark=-, mark size=3, mark options={solid}, only marks, forget plot]
table {%
0 1
1 1.015
2 0.983
3 0.978
4 0.964
5 3.503
6 9.94
};
\path [draw=color1, line width=0.04pt]
(axis cs:0,0.99)
--(axis cs:0,0.99);

\path [draw=color1, line width=0.04pt]
(axis cs:1,0.968)
--(axis cs:1,0.974);

\path [draw=color1, line width=0.04pt]
(axis cs:2,0.919)
--(axis cs:2,0.937);

\path [draw=color1, line width=0.04pt]
(axis cs:3,0.791)
--(axis cs:3,0.867);

\path [draw=color1, line width=0.04pt]
(axis cs:4,0.402)
--(axis cs:4,0.666);

\path [draw=color1, line width=0.04pt]
(axis cs:5,0.165)
--(axis cs:5,0.733);

\path [draw=color1, line width=0.04pt]
(axis cs:6,0.279)
--(axis cs:6,1.949);

\addplot [very thick, color1, mark=-, mark size=3, mark options={solid}, only marks, forget plot]
table {%
0 0.99
1 0.968
2 0.919
3 0.791
4 0.402
5 0.165
6 0.279
};
\addplot [very thick, color1, mark=-, mark size=3, mark options={solid}, only marks, forget plot]
table {%
0 0.99
1 0.974
2 0.937
3 0.867
4 0.666
5 0.733
6 1.949
};
\path [draw=color2, line width=0.04pt]
(axis cs:0,0.99)
--(axis cs:0,0.99);

\path [draw=color2, line width=0.04pt]
(axis cs:1,0.97)
--(axis cs:1,0.97);

\path [draw=color2, line width=0.04pt]
(axis cs:2,0.93)
--(axis cs:2,0.93);

\path [draw=color2, line width=0.04pt]
(axis cs:3,0.805)
--(axis cs:3,0.817);

\path [draw=color2, line width=0.04pt]
(axis cs:4,0.589)
--(axis cs:4,0.619);

\path [draw=color2, line width=0.04pt]
(axis cs:5,0.383)
--(axis cs:5,0.439);

\path [draw=color2, line width=0.04pt]
(axis cs:6,0.032)
--(axis cs:6,0.244);

\addplot [very thick, color2, mark=-, mark size=3, mark options={solid}, only marks, forget plot]
table {%
0 0.99
1 0.97
2 0.93
3 0.805
4 0.589
5 0.383
6 0.032
};
\addplot [very thick, color2, mark=-, mark size=3, mark options={solid}, only marks, forget plot]
table {%
0 0.99
1 0.97
2 0.93
3 0.817
4 0.619
5 0.439
6 0.244
};
\path [draw=color3, line width=0.04pt]
(axis cs:0,0.99)
--(axis cs:0,0.99);

\path [draw=color3, line width=0.04pt]
(axis cs:1,0.97)
--(axis cs:1,0.97);

\path [draw=color3, line width=0.04pt]
(axis cs:2,0.93)
--(axis cs:2,0.93);

\path [draw=color3, line width=0.04pt]
(axis cs:3,0.81)
--(axis cs:3,0.81);

\path [draw=color3, line width=0.04pt]
(axis cs:4,0.6)
--(axis cs:4,0.6);

\path [draw=color3, line width=0.04pt]
(axis cs:5,0.398)
--(axis cs:5,0.406);

\path [draw=color3, line width=0.04pt]
(axis cs:6,0.002)
--(axis cs:6,0.032);

\addplot [very thick, color3, mark=-, mark size=3, mark options={solid}, only marks, forget plot]
table {%
0 0.99
1 0.97
2 0.93
3 0.81
4 0.6
5 0.398
6 0.002
};
\addplot [very thick, color3, mark=-, mark size=3, mark options={solid}, only marks, forget plot]
table {%
0 0.99
1 0.97
2 0.93
3 0.81
4 0.6
5 0.406
6 0.032
};
\path [draw=color4, line width=0.04pt]
(axis cs:0,0.99)
--(axis cs:0,0.99);

\path [draw=color4, line width=0.04pt]
(axis cs:1,0.97)
--(axis cs:1,0.97);

\path [draw=color4, line width=0.04pt]
(axis cs:2,0.93)
--(axis cs:2,0.93);

\path [draw=color4, line width=0.04pt]
(axis cs:3,0.81)
--(axis cs:3,0.81);

\path [draw=color4, line width=0.04pt]
(axis cs:4,0.6)
--(axis cs:4,0.6);

\path [draw=color4, line width=0.04pt]
(axis cs:5,0.4)
--(axis cs:5,0.4);

\path [draw=color4, line width=0.04pt]
(axis cs:6,-0.002)
--(axis cs:6,0.004);

\addplot [very thick, color4, mark=-, mark size=3, mark options={solid}, only marks, forget plot]
table {%
0 0.99
1 0.97
2 0.93
3 0.81
4 0.6
5 0.4
6 -0.002
};
\addplot [very thick, color4, mark=-, mark size=3, mark options={solid}, only marks, forget plot]
table {%
0 0.99
1 0.97
2 0.93
3 0.81
4 0.6
5 0.4
6 0.004
};
\path [draw=white!27.843137254902!black, line width=0.04pt]
(axis cs:0,0.99)
--(axis cs:0,0.99);

\path [draw=white!27.843137254902!black, line width=0.04pt]
(axis cs:1,0.97)
--(axis cs:1,0.97);

\path [draw=white!27.843137254902!black, line width=0.04pt]
(axis cs:2,0.93)
--(axis cs:2,0.93);

\path [draw=white!27.843137254902!black, line width=0.04pt]
(axis cs:3,0.81)
--(axis cs:3,0.81);

\path [draw=white!27.843137254902!black, line width=0.04pt]
(axis cs:4,0.6)
--(axis cs:4,0.6);

\path [draw=white!27.843137254902!black, line width=0.04pt]
(axis cs:5,0.4)
--(axis cs:5,0.4);

\path [draw=white!27.843137254902!black, line width=0.04pt]
(axis cs:6,0)
--(axis cs:6,0);

\addplot [very thick, white!27.843137254902!black, mark=-, mark size=3, mark options={solid}, only marks, forget plot]
table {%
0 0.99
1 0.97
2 0.93
3 0.81
4 0.6
5 0.4
6 0
};
\addplot [very thick, white!27.843137254902!black, mark=-, mark size=3, mark options={solid}, only marks, forget plot]
table {%
0 0.99
1 0.97
2 0.93
3 0.81
4 0.6
5 0.4
6 0
};
\addplot [very thick, color0, dash pattern=on 1pt off 2pt]
table {%
0 0.986
1 0.978
2 0.896
3 0.742
4 0.621
5 1.743
6 4.543
};
\addplot [very thick, color1, dash pattern=on 3pt off 1pt on 1pt off 1pt on 1pt off 1pt]
table {%
0 0.99
1 0.971
2 0.928
3 0.829
4 0.534
5 0.449
6 1.114
};
\addplot [very thick, color2, dash pattern=on 5pt off 1pt]
table {%
0 0.99
1 0.97
2 0.93
3 0.811
4 0.604
5 0.411
6 0.138
};
\addplot [very thick, color3, dash pattern=on 1pt off 3pt on 3pt off 3pt]
table {%
0 0.99
1 0.97
2 0.93
3 0.81
4 0.6
5 0.402
6 0.017
};
\addplot [very thick, color4, dash pattern=on 1pt off 1pt]
table {%
0 0.99
1 0.97
2 0.93
3 0.81
4 0.6
5 0.4
6 0.001
};
\addplot [very thick, white!27.843137254902!black]
table {%
0 0.99
1 0.97
2 0.93
3 0.81
4 0.6
5 0.4
6 0
};
\addplot [very thick, color0, dash pattern=on 1pt off 2pt, mark=square*, mark size=0.5, mark options={solid}, forget plot]
table {%
0 0.986
1 0.978
2 0.896
3 0.742
4 0.621
5 1.743
6 4.543
};
\addplot [very thick, color1, dash pattern=on 3pt off 1pt on 1pt off 1pt on 1pt off 1pt, mark=square*, mark size=0.5, mark options={solid}, forget plot]
table {%
0 0.99
1 0.971
2 0.928
3 0.829
4 0.534
5 0.449
6 1.114
};
\addplot [very thick, color2, dash pattern=on 5pt off 1pt, mark=square*, mark size=0.5, mark options={solid}, forget plot]
table {%
0 0.99
1 0.97
2 0.93
3 0.811
4 0.604
5 0.411
6 0.138
};
\addplot [very thick, color3, dash pattern=on 1pt off 3pt on 3pt off 3pt, mark=square*, mark size=0.5, mark options={solid}, forget plot]
table {%
0 0.99
1 0.97
2 0.93
3 0.81
4 0.6
5 0.402
6 0.017
};
\addplot [very thick, color4, dash pattern=on 1pt off 1pt, mark=square*, mark size=0.5, mark options={solid}, forget plot]
table {%
0 0.99
1 0.97
2 0.93
3 0.81
4 0.6
5 0.4
6 0.001
};
\addplot [very thick, white!27.843137254902!black, mark=square*, mark size=0.5, mark options={solid}, forget plot]
table {%
0 0.99
1 0.97
2 0.93
3 0.81
4 0.6
5 0.4
6 0
};
\end{axis}

\end{tikzpicture}
	\end{minipage}
	\hspace{2em}
	\begin{minipage}[t]{0.3\textwidth}
\begin{tikzpicture}[scale=0.6]

\definecolor{color0}{rgb}{0.0470588235294118,0.364705882352941,0.647058823529412}
\definecolor{color1}{rgb}{0,0.725490196078431,0.270588235294118}
\definecolor{color2}{rgb}{1,0.584313725490196,0}
\definecolor{color3}{rgb}{1,0.172549019607843,0}
\definecolor{color4}{rgb}{0.717647058823529,0.454901960784314,0.83921568627451}

\begin{axis}[
scaled y ticks=manual:{}{\pgfmathparse{#1}},
tick pos=both,
x grid style={white!69.0196078431373!black},
xlabel={M},
xmin=-0.2, xmax=4.2,
xtick style={color=black},
xtick={0,1,2,3,4},
xtick={0,1,2,3,4},
xtick={0,1,2,3,4},
xtick={0,1,2,3,4},
xtick={0,1,2,3,4},
xtick={0,1,2,3,4},
xtick={0,1,2,3,4},
xtick={0,1,2,3,4},
xtick={0,1,2,3,4},
xtick={0,1,2,3,4},
xtick={0,1,2,3,4},
xtick={0,1,2,3,4},
xtick={0,1,2,3,4},
xtick={0,1,2,3,4},
xtick={0,1,2,3,4},
xtick={0,1,2,3,4},
xtick={0,1,2,3,4},
xtick={0,1,2,3,4},
xticklabels={1,2,4,5,20},
xticklabels={1,2,4,5,20},
xticklabels={1,2,4,5,20},
xticklabels={1,2,4,5,20},
xticklabels={1,2,4,5,20},
xticklabels={1,2,4,5,20},
xticklabels={1,2,4,5,20},
xticklabels={1,2,4,5,20},
xticklabels={1,2,4,5,20},
xticklabels={1,2,4,5,20},
xticklabels={1,2,4,5,20},
xticklabels={1,2,4,5,20},
xticklabels={1,2,4,5,20},
xticklabels={1,2,4,5,20},
xticklabels={1,2,4,5,20},
xticklabels={1,2,4,5,20},
xticklabels={1,2,4,5,20},
xticklabels={1,2,4,5,20},
y grid style={white!69.0196078431373!black},
ymin=-1, ymax=5,
ytick style={color=black},
]
\path [draw=color0, line width=0.04pt]
(axis cs:0,0.66)
--(axis cs:0,0.66);

\path [draw=color0, line width=0.04pt]
(axis cs:1,0.331)
--(axis cs:1,0.341);

\path [draw=color0, line width=0.04pt]
(axis cs:2,0.073)
--(axis cs:2,0.087);

\path [draw=color0, line width=0.04pt]
(axis cs:3,0.04)
--(axis cs:3,0.05);

\path [draw=color0, line width=0.04pt]
(axis cs:4,-0.004)
--(axis cs:4,0.028);

\addplot [very thick, color0, mark=-, mark size=3, mark options={solid}, only marks]
table {%
0 0.66
1 0.331
2 0.073
3 0.04
4 -0.004
};
\addplot [very thick, color0, mark=-, mark size=3, mark options={solid}, only marks]
table {%
0 0.66
1 0.341
2 0.087
3 0.05
4 0.028
};
\path [draw=color1, line width=0.04pt]
(axis cs:0,0.66)
--(axis cs:0,0.66);

\path [draw=color1, line width=0.04pt]
(axis cs:1,0.33)
--(axis cs:1,0.33);

\path [draw=color1, line width=0.04pt]
(axis cs:2,0.08)
--(axis cs:2,0.08);

\path [draw=color1, line width=0.04pt]
(axis cs:3,0.04)
--(axis cs:3,0.04);

\path [draw=color1, line width=0.04pt]
(axis cs:4,0)
--(axis cs:4,0);

\addplot [very thick, color1, mark=-, mark size=3, mark options={solid}, only marks]
table {%
0 0.66
1 0.33
2 0.08
3 0.04
4 0
};
\addplot [very thick, color1, mark=-, mark size=3, mark options={solid}, only marks]
table {%
0 0.66
1 0.33
2 0.08
3 0.04
4 0
};
\path [draw=color2, line width=0.04pt]
(axis cs:0,0.66)
--(axis cs:0,0.66);

\path [draw=color2, line width=0.04pt]
(axis cs:1,0.33)
--(axis cs:1,0.33);

\path [draw=color2, line width=0.04pt]
(axis cs:2,0.08)
--(axis cs:2,0.08);

\path [draw=color2, line width=0.04pt]
(axis cs:3,0.04)
--(axis cs:3,0.04);

\path [draw=color2, line width=0.04pt]
(axis cs:4,0)
--(axis cs:4,0);

\addplot [very thick, color2, mark=-, mark size=3, mark options={solid}, only marks]
table {%
0 0.66
1 0.33
2 0.08
3 0.04
4 0
};
\addplot [very thick, color2, mark=-, mark size=3, mark options={solid}, only marks]
table {%
0 0.66
1 0.33
2 0.08
3 0.04
4 0
};
\path [draw=color3, line width=0.04pt]
(axis cs:0,0.66)
--(axis cs:0,0.66);

\path [draw=color3, line width=0.04pt]
(axis cs:1,0.33)
--(axis cs:1,0.33);

\path [draw=color3, line width=0.04pt]
(axis cs:2,0.08)
--(axis cs:2,0.08);

\path [draw=color3, line width=0.04pt]
(axis cs:3,0.04)
--(axis cs:3,0.04);

\path [draw=color3, line width=0.04pt]
(axis cs:4,0)
--(axis cs:4,0);

\addplot [very thick, color3, mark=-, mark size=3, mark options={solid}, only marks]
table {%
0 0.66
1 0.33
2 0.08
3 0.04
4 0
};
\addplot [very thick, color3, mark=-, mark size=3, mark options={solid}, only marks]
table {%
0 0.66
1 0.33
2 0.08
3 0.04
4 0
};
\path [draw=color4, line width=0.04pt]
(axis cs:0,0.66)
--(axis cs:0,0.66);

\path [draw=color4, line width=0.04pt]
(axis cs:1,0.33)
--(axis cs:1,0.33);

\path [draw=color4, line width=0.04pt]
(axis cs:2,0.08)
--(axis cs:2,0.08);

\path [draw=color4, line width=0.04pt]
(axis cs:3,0.04)
--(axis cs:3,0.04);

\path [draw=color4, line width=0.04pt]
(axis cs:4,0)
--(axis cs:4,0);

\addplot [very thick, color4, mark=-, mark size=3, mark options={solid}, only marks]
table {%
0 0.66
1 0.33
2 0.08
3 0.04
4 0
};
\addplot [very thick, color4, mark=-, mark size=3, mark options={solid}, only marks]
table {%
0 0.66
1 0.33
2 0.08
3 0.04
4 0
};
\path [draw=white!27.843137254902!black, line width=0.04pt]
(axis cs:0,0.66)
--(axis cs:0,0.66);

\path [draw=white!27.843137254902!black, line width=0.04pt]
(axis cs:1,0.33)
--(axis cs:1,0.33);

\path [draw=white!27.843137254902!black, line width=0.04pt]
(axis cs:2,0.08)
--(axis cs:2,0.08);

\path [draw=white!27.843137254902!black, line width=0.04pt]
(axis cs:3,0.04)
--(axis cs:3,0.04);

\path [draw=white!27.843137254902!black, line width=0.04pt]
(axis cs:4,0)
--(axis cs:4,0);

\addplot [very thick, white!27.843137254902!black, mark=-, mark size=3, mark options={solid}, only marks]
table {%
0 0.66
1 0.33
2 0.08
3 0.04
4 0
};
\addplot [very thick, white!27.843137254902!black, mark=-, mark size=3, mark options={solid}, only marks]
table {%
0 0.66
1 0.33
2 0.08
3 0.04
4 0
};
\addplot [very thick, color0, dash pattern=on 1pt off 2pt, mark=square*, mark size=0.5, mark options={solid}]
table {%
0 0.66
1 0.336
2 0.08
3 0.045
4 0.012
};
\addplot [very thick, color1, dash pattern=on 3pt off 1pt on 1pt off 1pt on 1pt off 1pt, mark=square*, mark size=0.5, mark options={solid}]
table {%
0 0.66
1 0.33
2 0.08
3 0.04
4 0
};
\addplot [very thick, color2, dash pattern=on 5pt off 1pt, mark=square*, mark size=0.5, mark options={solid}]
table {%
0 0.66
1 0.33
2 0.08
3 0.04
4 0
};
\addplot [very thick, color3, dash pattern=on 1pt off 3pt on 3pt off 3pt, mark=square*, mark size=0.5, mark options={solid}]
table {%
0 0.66
1 0.33
2 0.08
3 0.04
4 0
};
\addplot [very thick, color4, dash pattern=on 1pt off 1pt, mark=square*, mark size=0.5, mark options={solid}]
table {%
0 0.66
1 0.33
2 0.08
3 0.04
4 0
};
\addplot [very thick, white!27.843137254902!black, mark=square*, mark size=0.5, mark options={solid}]
table {%
0 0.66
1 0.33
2 0.08
3 0.04
4 0
};
\end{axis}

\end{tikzpicture}
	\end{minipage}
	\begin{minipage}[t]{0.3\textwidth}
\begin{tikzpicture}[scale=0.6]

\definecolor{color0}{rgb}{0.0470588235294118,0.364705882352941,0.647058823529412}
\definecolor{color1}{rgb}{0,0.725490196078431,0.270588235294118}
\definecolor{color2}{rgb}{1,0.584313725490196,0}
\definecolor{color3}{rgb}{1,0.172549019607843,0}
\definecolor{color4}{rgb}{0.717647058823529,0.454901960784314,0.83921568627451}

\begin{axis}[
scaled y ticks=manual:{}{\pgfmathparse{#1}},
tick pos=both,
x grid style={white!69.0196078431373!black},
xlabel={M},
xmin=-0.25, xmax=5.25,
xtick style={color=black},
xtick={0,1,2,3,4,5},
xtick={0,1,2,3,4,5},
xtick={0,1,2,3,4,5},
xtick={0,1,2,3,4,5},
xtick={0,1,2,3,4,5},
xtick={0,1,2,3,4,5},
xtick={0,1,2,3,4,5},
xtick={0,1,2,3,4,5},
xtick={0,1,2,3,4,5},
xtick={0,1,2,3,4,5},
xtick={0,1,2,3,4,5},
xtick={0,1,2,3,4,5},
xtick={0,1,2,3,4,5},
xtick={0,1,2,3,4,5},
xtick={0,1,2,3,4,5},
xtick={0,1,2,3,4,5},
xtick={0,1,2,3,4,5},
xtick={0,1,2,3,4,5},
xticklabels={1,2,4,5,6,16},
xticklabels={1,2,4,5,6,16},
xticklabels={1,2,4,5,6,16},
xticklabels={1,2,4,5,6,16},
xticklabels={1,2,4,5,6,16},
xticklabels={1,2,4,5,6,16},
xticklabels={1,2,4,5,6,16},
xticklabels={1,2,4,5,6,16},
xticklabels={1,2,4,5,6,16},
xticklabels={1,2,4,5,6,16},
xticklabels={1,2,4,5,6,16},
xticklabels={1,2,4,5,6,16},
xticklabels={1,2,4,5,6,16},
xticklabels={1,2,4,5,6,16},
xticklabels={1,2,4,5,6,16},
xticklabels={1,2,4,5,6,16},
xticklabels={1,2,4,5,6,16},
xticklabels={1,2,4,5,6,16},
y grid style={white!69.0196078431373!black},
ymin=-1, ymax=5,
ytick style={color=black},
]
\path [draw=color0, line width=0.04pt]
(axis cs:0,0.73)
--(axis cs:0,0.73);

\path [draw=color0, line width=0.04pt]
(axis cs:1,0.464)
--(axis cs:1,0.472);

\path [draw=color0, line width=0.04pt]
(axis cs:2,0.148)
--(axis cs:2,0.154);

\path [draw=color0, line width=0.04pt]
(axis cs:3,0.078)
--(axis cs:3,0.088);

\path [draw=color0, line width=0.04pt]
(axis cs:4,0.04)
--(axis cs:4,0.04);

\path [draw=color0, line width=0.04pt]
(axis cs:5,-0.002)
--(axis cs:5,0.004);

\addplot [very thick, color0, mark=-, mark size=3, mark options={solid}, only marks]
table {%
0 0.73
1 0.464
2 0.148
3 0.078
4 0.04
5 -0.002
};
\addplot [very thick, color0, mark=-, mark size=3, mark options={solid}, only marks]
table {%
0 0.73
1 0.472
2 0.154
3 0.088
4 0.04
5 0.004
};
\path [draw=color1, line width=0.04pt]
(axis cs:0,0.73)
--(axis cs:0,0.73);

\path [draw=color1, line width=0.04pt]
(axis cs:1,0.47)
--(axis cs:1,0.47);

\path [draw=color1, line width=0.04pt]
(axis cs:2,0.15)
--(axis cs:2,0.15);

\path [draw=color1, line width=0.04pt]
(axis cs:3,0.08)
--(axis cs:3,0.08);

\path [draw=color1, line width=0.04pt]
(axis cs:4,0.04)
--(axis cs:4,0.04);

\path [draw=color1, line width=0.04pt]
(axis cs:5,0)
--(axis cs:5,0);

\addplot [very thick, color1, mark=-, mark size=3, mark options={solid}, only marks]
table {%
0 0.73
1 0.47
2 0.15
3 0.08
4 0.04
5 0
};
\addplot [very thick, color1, mark=-, mark size=3, mark options={solid}, only marks]
table {%
0 0.73
1 0.47
2 0.15
3 0.08
4 0.04
5 0
};
\path [draw=color2, line width=0.04pt]
(axis cs:0,0.73)
--(axis cs:0,0.73);

\path [draw=color2, line width=0.04pt]
(axis cs:1,0.47)
--(axis cs:1,0.47);

\path [draw=color2, line width=0.04pt]
(axis cs:2,0.15)
--(axis cs:2,0.15);

\path [draw=color2, line width=0.04pt]
(axis cs:3,0.08)
--(axis cs:3,0.08);

\path [draw=color2, line width=0.04pt]
(axis cs:4,0.04)
--(axis cs:4,0.04);

\path [draw=color2, line width=0.04pt]
(axis cs:5,0)
--(axis cs:5,0);

\addplot [very thick, color2, mark=-, mark size=3, mark options={solid}, only marks]
table {%
0 0.73
1 0.47
2 0.15
3 0.08
4 0.04
5 0
};
\addplot [very thick, color2, mark=-, mark size=3, mark options={solid}, only marks]
table {%
0 0.73
1 0.47
2 0.15
3 0.08
4 0.04
5 0
};
\path [draw=color3, line width=0.04pt]
(axis cs:0,0.73)
--(axis cs:0,0.73);

\path [draw=color3, line width=0.04pt]
(axis cs:1,0.47)
--(axis cs:1,0.47);

\path [draw=color3, line width=0.04pt]
(axis cs:2,0.15)
--(axis cs:2,0.15);

\path [draw=color3, line width=0.04pt]
(axis cs:3,0.08)
--(axis cs:3,0.08);

\path [draw=color3, line width=0.04pt]
(axis cs:4,0.04)
--(axis cs:4,0.04);

\path [draw=color3, line width=0.04pt]
(axis cs:5,0)
--(axis cs:5,0);

\addplot [very thick, color3, mark=-, mark size=3, mark options={solid}, only marks]
table {%
0 0.73
1 0.47
2 0.15
3 0.08
4 0.04
5 0
};
\addplot [very thick, color3, mark=-, mark size=3, mark options={solid}, only marks]
table {%
0 0.73
1 0.47
2 0.15
3 0.08
4 0.04
5 0
};
\path [draw=color4, line width=0.04pt]
(axis cs:0,0.73)
--(axis cs:0,0.73);

\path [draw=color4, line width=0.04pt]
(axis cs:1,0.47)
--(axis cs:1,0.47);

\path [draw=color4, line width=0.04pt]
(axis cs:2,0.15)
--(axis cs:2,0.15);

\path [draw=color4, line width=0.04pt]
(axis cs:3,0.08)
--(axis cs:3,0.08);

\path [draw=color4, line width=0.04pt]
(axis cs:4,0.04)
--(axis cs:4,0.04);

\path [draw=color4, line width=0.04pt]
(axis cs:5,0)
--(axis cs:5,0);

\addplot [very thick, color4, mark=-, mark size=3, mark options={solid}, only marks]
table {%
0 0.73
1 0.47
2 0.15
3 0.08
4 0.04
5 0
};
\addplot [very thick, color4, mark=-, mark size=3, mark options={solid}, only marks]
table {%
0 0.73
1 0.47
2 0.15
3 0.08
4 0.04
5 0
};
\path [draw=white!27.843137254902!black, line width=0.04pt]
(axis cs:0,0.73)
--(axis cs:0,0.73);

\path [draw=white!27.843137254902!black, line width=0.04pt]
(axis cs:1,0.47)
--(axis cs:1,0.47);

\path [draw=white!27.843137254902!black, line width=0.04pt]
(axis cs:2,0.15)
--(axis cs:2,0.15);

\path [draw=white!27.843137254902!black, line width=0.04pt]
(axis cs:3,0.08)
--(axis cs:3,0.08);

\path [draw=white!27.843137254902!black, line width=0.04pt]
(axis cs:4,0.04)
--(axis cs:4,0.04);

\path [draw=white!27.843137254902!black, line width=0.04pt]
(axis cs:5,0)
--(axis cs:5,0);

\addplot [very thick, white!27.843137254902!black, mark=-, mark size=3, mark options={solid}, only marks]
table {%
0 0.73
1 0.47
2 0.15
3 0.08
4 0.04
5 0
};
\addplot [very thick, white!27.843137254902!black, mark=-, mark size=3, mark options={solid}, only marks]
table {%
0 0.73
1 0.47
2 0.15
3 0.08
4 0.04
5 0
};
\addplot [very thick, color0, dash pattern=on 1pt off 2pt, mark=square*, mark size=0.5, mark options={solid}]
table {%
0 0.73
1 0.468
2 0.151
3 0.083
4 0.04
5 0.001
};
\addplot [very thick, color1, dash pattern=on 3pt off 1pt on 1pt off 1pt on 1pt off 1pt, mark=square*, mark size=0.5, mark options={solid}]
table {%
0 0.73
1 0.47
2 0.15
3 0.08
4 0.04
5 0
};
\addplot [very thick, color2, dash pattern=on 5pt off 1pt, mark=square*, mark size=0.5, mark options={solid}]
table {%
0 0.73
1 0.47
2 0.15
3 0.08
4 0.04
5 0
};
\addplot [very thick, color3, dash pattern=on 1pt off 3pt on 3pt off 3pt, mark=square*, mark size=0.5, mark options={solid}]
table {%
0 0.73
1 0.47
2 0.15
3 0.08
4 0.04
5 0
};
\addplot [very thick, color4, dash pattern=on 1pt off 1pt, mark=square*, mark size=0.5, mark options={solid}]
table {%
0 0.73
1 0.47
2 0.15
3 0.08
4 0.04
5 0
};
\addplot [very thick, white!27.843137254902!black, mark=square*, mark size=0.5, mark options={solid}]
table {%
0 0.73
1 0.47
2 0.15
3 0.08
4 0.04
5 0
};
\end{axis}

\end{tikzpicture}
	\end{minipage}
	\begin{minipage}[t]{\textwidth}
		\vspace{-1em}
		\centering
\begin{tikzpicture}[scale=0.6]
	
	\definecolor{color0}{rgb}{0.0470588235294118,0.364705882352941,0.647058823529412}
	\definecolor{color1}{rgb}{0,0.725490196078431,0.270588235294118}
	\definecolor{color2}{rgb}{1,0.584313725490196,0}
	\definecolor{color3}{rgb}{1,0.172549019607843,0}
	\definecolor{color4}{rgb}{0.717647058823529,0.454901960784314,0.83921568627451}

	\begin{axis}[
		hide axis,
		xmin=0,
		xmax=1,
		ymin=0,
		ymax=1,		
		legend cell align={left},
		legend columns=7,
		legend style={
			fill opacity=0.8,
			draw opacity=1,
			text opacity=1,
			draw=white!80!black
		}
		]
		\addlegendimage{empty legend}
		\addlegendentry{$\varepsilon$: }
		\addlegendimage{very thick, color0, dash pattern=on 1pt off 2pt}
		\addlegendentry{0.01}
		\addlegendimage{very thick, color1, dash pattern=on 3pt off 1pt on 1pt 
		off 1pt on 1pt off 1pt}
		\addlegendentry{0.1}
		\addlegendimage{very thick, color2, dash pattern=on 5pt off 1pt}
		\addlegendentry{1}
		\addlegendimage{very thick, color3, dash pattern=on 1pt off 3pt on 3pt 
		off 3pt}
		\addlegendentry{10}
		\addlegendimage{very thick, color4, dash pattern=on 1pt off 1pt}
		\addlegendentry{100}
		\addlegendimage{very thick, white!27.843137254902!black}
		\addlegendentry{withoutDp}
		\addlegendimage{very thick, color0, dash pattern=on 1pt off 2pt, 
		mark=square*, mark size=0.5, mark options={solid}, forget plot}

	\end{axis}

\end{tikzpicture}
	\end{minipage}
	\begin{minipage}[t]{\textwidth}
		\captionsetup[sub]{labelformat=parens}
		\begingroup
		\captionsetup{type=figure}
		\begin{subfigure}{0.3\textwidth}
			\vspace*{-5cm}
			\caption{\tt GEOLIFE}
		\end{subfigure}
		\begin{subfigure}{0.3\textwidth}
			\vspace*{-5cm}
			\caption{\tt MADRID}
		\end{subfigure}
		\begin{subfigure}{0.3\textwidth}
			\vspace*{-5cm}
			\caption{\tt BERLIN}
		\end{subfigure}
		\endgroup
	\end{minipage}
	\begin{minipage}[t]{\textwidth}
		\vspace*{-2.7cm}
		\caption{$\mathrm{TripCountError}$ for different values of $M$
			and $\varepsilon$.
			The standard deviation of all 10 runs is represented by the error 
			bars.}
		\label{fig:tripError}
	\end{minipage}
\vspace{-2.5cm}
\end{figure}

Without the application of privacy measures, sampling of data decreases the similarity to its origin. 
However, a sampling that bounds the user contributions to a defined maximum $M$ can increase the 
similarity 
when user-level differential privacy guarantees are enforced. This is because $M$ defines the sensitivity 
which in turn defines the amount of added noise. 
Thus, the question arises which effect outweighs the other: the increase of similarity by 
keeping more data points per user or the decrease of noise by reducing $M$. The effect highly depends 
on the considered mobility measure and data set which we will evaluate in depth 
in the following. Note that 
the errors produced purely by sampling without any additional noise, denoted 
\textit{withoutDp},
function as a lower bound for the differentially private error values.

The evaluation focuses on three aspects: (1) the overall range of error values, (2) the deviation 
from the baseline \textit{withoutDp}, and (3) the existence of `tipping points' for error values created 
through the interplay of $M$ and $\varepsilon$.

As expected, the effect of additional information dominates that of
additional noise for almost all data sets and $\varepsilon$-values for the $\mathrm{TripCountError}$ (see 
Figure~\ref{fig:tripError}). 
Almost all $\varepsilon$-curves resemble the baseline \textit{withoutDp}, i.e. 
the added noise has no influence on the error. Instead, the error can be 
attributed to the 
information loss through sampling. $M = 1$ yields a high relative error between 
66\,\% and 100\,\% for all 
three data sets. An absolute count, unlike a relative share, intuitively largely gains similarity to its origin with 
an increased sample. Accordingly, the curves drop for an increase in $M$. 
But even for such a top-level aggregation, a lower sensitivity can outweigh the 
information loss: for e.g. $\varepsilon = 0.1$, the error for the highly skewed data set {\tt GEOLIFE} 
increases 
from 32\,\% to 103\,\% when setting $M = 2,153$ instead of $M = 215$.
More fine granular mobility measures where trip counts are disaggregated, e.g., 
geospatially into tiles, 
yield smaller values which are in turn more sensitive to adding noise. 

\begin{figure}[tb]
	\centering
	\begin{minipage}[t]{0.3\textwidth}	
		\input{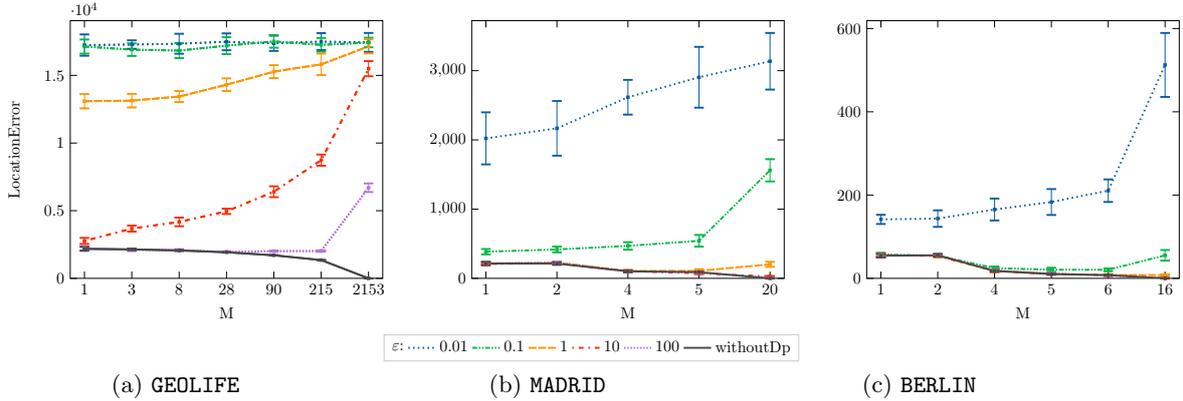}
	\end{minipage}
	\hspace{1.5em}
	\begin{minipage}[t]{0.3\textwidth}
\begin{tikzpicture}[scale=0.6]

\definecolor{color0}{rgb}{0.0470588235294118,0.364705882352941,0.647058823529412}
\definecolor{color1}{rgb}{0,0.725490196078431,0.270588235294118}
\definecolor{color2}{rgb}{1,0.584313725490196,0}
\definecolor{color3}{rgb}{1,0.172549019607843,0}
\definecolor{color4}{rgb}{0.717647058823529,0.454901960784314,0.83921568627451}

\begin{axis}[
tick pos=both,
x grid style={white!69.0196078431373!black},
xlabel={M},
xmin=-0.2, xmax=4.2,
xtick style={color=black},
xtick={0,1,2,3,4},
xtick={0,1,2,3,4},
xtick={0,1,2,3,4},
xtick={0,1,2,3,4},
xtick={0,1,2,3,4},
xtick={0,1,2,3,4},
xtick={0,1,2,3,4},
xtick={0,1,2,3,4},
xtick={0,1,2,3,4},
xtick={0,1,2,3,4},
xtick={0,1,2,3,4},
xtick={0,1,2,3,4},
xtick={0,1,2,3,4},
xtick={0,1,2,3,4},
xtick={0,1,2,3,4},
xtick={0,1,2,3,4},
xtick={0,1,2,3,4},
xtick={0,1,2,3,4},
xticklabels={1,2,4,5,20},
xticklabels={1,2,4,5,20},
xticklabels={1,2,4,5,20},
xticklabels={1,2,4,5,20},
xticklabels={1,2,4,5,20},
xticklabels={1,2,4,5,20},
xticklabels={1,2,4,5,20},
xticklabels={1,2,4,5,20},
xticklabels={1,2,4,5,20},
xticklabels={1,2,4,5,20},
xticklabels={1,2,4,5,20},
xticklabels={1,2,4,5,20},
xticklabels={1,2,4,5,20},
xticklabels={1,2,4,5,20},
xticklabels={1,2,4,5,20},
xticklabels={1,2,4,5,20},
xticklabels={1,2,4,5,20},
xticklabels={1,2,4,5,20},
y grid style={white!69.0196078431373!black},
ymin=0, ymax=3722.2668,
ytick style={color=black}
]
\path [draw=color0, line width=0.04pt]
(axis cs:0,1646.384)
--(axis cs:0,2398.372);

\path [draw=color0, line width=0.04pt]
(axis cs:1,1772.062)
--(axis cs:1,2561.744);

\path [draw=color0, line width=0.04pt]
(axis cs:2,2365.273)
--(axis cs:2,2867.327);

\path [draw=color0, line width=0.04pt]
(axis cs:3,2465.581)
--(axis cs:3,3344.005);

\path [draw=color0, line width=0.04pt]
(axis cs:4,2727.104)
--(axis cs:4,3545.016);

\addplot [very thick, color0, mark=-, mark size=3, mark options={solid}, only marks]
table {%
0 1646.384
1 1772.062
2 2365.273
3 2465.581
4 2727.104
};
\addplot [very thick, color0, mark=-, mark size=3, mark options={solid}, only marks]
table {%
0 2398.372
1 2561.744
2 2867.327
3 3344.005
4 3545.016
};
\path [draw=color1, line width=0.04pt]
(axis cs:0,346.18)
--(axis cs:0,423.864);

\path [draw=color1, line width=0.04pt]
(axis cs:1,375.899)
--(axis cs:1,459.199);

\path [draw=color1, line width=0.04pt]
(axis cs:2,415.025)
--(axis cs:2,522.633);

\path [draw=color1, line width=0.04pt]
(axis cs:3,458.024)
--(axis cs:3,626.946);

\path [draw=color1, line width=0.04pt]
(axis cs:4,1400.086)
--(axis cs:4,1721.73);

\addplot [very thick, color1, mark=-, mark size=3, mark options={solid}, only marks]
table {%
0 346.18
1 375.899
2 415.025
3 458.024
4 1400.086
};
\addplot [very thick, color1, mark=-, mark size=3, mark options={solid}, only marks]
table {%
0 423.864
1 459.199
2 522.633
3 626.946
4 1721.73
};
\path [draw=color2, line width=0.04pt]
(axis cs:0,193.137)
--(axis cs:0,233.601);

\path [draw=color2, line width=0.04pt]
(axis cs:1,212.467)
--(axis cs:1,240.719);

\path [draw=color2, line width=0.04pt]
(axis cs:2,94.373)
--(axis cs:2,120.819);

\path [draw=color2, line width=0.04pt]
(axis cs:3,84.574)
--(axis cs:3,132.154);

\path [draw=color2, line width=0.04pt]
(axis cs:4,165.7)
--(axis cs:4,238.588);

\addplot [very thick, color2, mark=-, mark size=3, mark options={solid}, only marks]
table {%
0 193.137
1 212.467
2 94.373
3 84.574
4 165.7
};
\addplot [very thick, color2, mark=-, mark size=3, mark options={solid}, only marks]
table {%
0 233.601
1 240.719
2 120.819
3 132.154
4 238.588
};
\path [draw=color3, line width=0.04pt]
(axis cs:0,190.702)
--(axis cs:0,224.418);

\path [draw=color3, line width=0.04pt]
(axis cs:1,202.788)
--(axis cs:1,232.994);

\path [draw=color3, line width=0.04pt]
(axis cs:2,90.543)
--(axis cs:2,116.925);

\path [draw=color3, line width=0.04pt]
(axis cs:3,59.534)
--(axis cs:3,84.588);

\path [draw=color3, line width=0.04pt]
(axis cs:4,23.245)
--(axis cs:4,33.683);

\addplot [very thick, color3, mark=-, mark size=3, mark options={solid}, only marks]
table {%
0 190.702
1 202.788
2 90.543
3 59.534
4 23.245
};
\addplot [very thick, color3, mark=-, mark size=3, mark options={solid}, only marks]
table {%
0 224.418
1 232.994
2 116.925
3 84.588
4 33.683
};
\path [draw=color4, line width=0.04pt]
(axis cs:0,194.577)
--(axis cs:0,239.859);

\path [draw=color4, line width=0.04pt]
(axis cs:1,216.606)
--(axis cs:1,243.57);

\path [draw=color4, line width=0.04pt]
(axis cs:2,90.86)
--(axis cs:2,111.22);

\path [draw=color4, line width=0.04pt]
(axis cs:3,65.982)
--(axis cs:3,102.046);

\path [draw=color4, line width=0.04pt]
(axis cs:4,1.221)
--(axis cs:4,7.483);

\addplot [very thick, color4, mark=-, mark size=3, mark options={solid}, only marks]
table {%
0 194.577
1 216.606
2 90.86
3 65.982
4 1.221
};
\addplot [very thick, color4, mark=-, mark size=3, mark options={solid}, only marks]
table {%
0 239.859
1 243.57
2 111.22
3 102.046
4 7.483
};
\path [draw=white!27.843137254902!black, line width=0.04pt]
(axis cs:0,191.711)
--(axis cs:0,238.397);

\path [draw=white!27.843137254902!black, line width=0.04pt]
(axis cs:1,194.314)
--(axis cs:1,231.038);

\path [draw=white!27.843137254902!black, line width=0.04pt]
(axis cs:2,87.541)
--(axis cs:2,115.543);

\path [draw=white!27.843137254902!black, line width=0.04pt]
(axis cs:3,81.552)
--(axis cs:3,97.536);

\path [draw=white!27.843137254902!black, line width=0.04pt]
(axis cs:4,0)
--(axis cs:4,0);

\addplot [very thick, white!27.843137254902!black, mark=-, mark size=3, mark options={solid}, only marks]
table {%
0 191.711
1 194.314
2 87.541
3 81.552
4 0
};
\addplot [very thick, white!27.843137254902!black, mark=-, mark size=3, mark options={solid}, only marks]
table {%
0 238.397
1 231.038
2 115.543
3 97.536
4 0
};
\addplot [very thick, color0, dash pattern=on 1pt off 2pt, mark=square*, mark size=0.5, mark options={solid}]
table {%
0 2022.378
1 2166.903
2 2616.3
3 2904.793
4 3136.06
};
\addplot [very thick, color1, dash pattern=on 3pt off 1pt on 1pt off 1pt on 1pt off 1pt, mark=square*, mark size=0.5, mark options={solid}]
table {%
0 385.022
1 417.549
2 468.829
3 542.485
4 1560.908
};
\addplot [very thick, color2, dash pattern=on 5pt off 1pt, mark=square*, mark size=0.5, mark options={solid}]
table {%
0 213.369
1 226.593
2 107.596
3 108.364
4 202.144
};
\addplot [very thick, color3, dash pattern=on 1pt off 3pt on 3pt off 3pt, mark=square*, mark size=0.5, mark options={solid}]
table {%
0 207.56
1 217.891
2 103.734
3 72.061
4 28.464
};
\addplot [very thick, color4, dash pattern=on 1pt off 1pt, mark=square*, mark size=0.5, mark options={solid}]
table {%
0 217.218
1 230.088
2 101.04
3 84.014
4 4.352
};
\addplot [very thick, white!27.843137254902!black, mark=square*, mark size=0.5, mark options={solid}]
table {%
0 215.054
1 212.676
2 101.542
3 89.544
4 0
};
\end{axis}

\end{tikzpicture}
	\end{minipage}
	\hspace{1em}
	\begin{minipage}[t]{0.3\textwidth}
\begin{tikzpicture}[scale=0.6]

\definecolor{color0}{rgb}{0.0470588235294118,0.364705882352941,0.647058823529412}
\definecolor{color1}{rgb}{0,0.725490196078431,0.270588235294118}
\definecolor{color2}{rgb}{1,0.584313725490196,0}
\definecolor{color3}{rgb}{1,0.172549019607843,0}
\definecolor{color4}{rgb}{0.717647058823529,0.454901960784314,0.83921568627451}

\begin{axis}[
tick pos=both,
x grid style={white!69.0196078431373!black},
xlabel={M},
xmin=-0.25, xmax=5.25,
xtick style={color=black},
xtick={0,1,2,3,4,5},
xtick={0,1,2,3,4,5},
xtick={0,1,2,3,4,5},
xtick={0,1,2,3,4,5},
xtick={0,1,2,3,4,5},
xtick={0,1,2,3,4,5},
xtick={0,1,2,3,4,5},
xtick={0,1,2,3,4,5},
xtick={0,1,2,3,4,5},
xtick={0,1,2,3,4,5},
xtick={0,1,2,3,4,5},
xtick={0,1,2,3,4,5},
xtick={0,1,2,3,4,5},
xtick={0,1,2,3,4,5},
xtick={0,1,2,3,4,5},
xtick={0,1,2,3,4,5},
xtick={0,1,2,3,4,5},
xtick={0,1,2,3,4,5},
xticklabels={1,2,4,5,6,16},
xticklabels={1,2,4,5,6,16},
xticklabels={1,2,4,5,6,16},
xticklabels={1,2,4,5,6,16},
xticklabels={1,2,4,5,6,16},
xticklabels={1,2,4,5,6,16},
xticklabels={1,2,4,5,6,16},
xticklabels={1,2,4,5,6,16},
xticklabels={1,2,4,5,6,16},
xticklabels={1,2,4,5,6,16},
xticklabels={1,2,4,5,6,16},
xticklabels={1,2,4,5,6,16},
xticklabels={1,2,4,5,6,16},
xticklabels={1,2,4,5,6,16},
xticklabels={1,2,4,5,6,16},
xticklabels={1,2,4,5,6,16},
xticklabels={1,2,4,5,6,16},
xticklabels={1,2,4,5,6,16},
y grid style={white!69.0196078431373!black},
ymin=0, ymax=618.79545,
ytick style={color=black}
]
\path [draw=color0, line width=0.04pt]
(axis cs:0,130.56)
--(axis cs:0,152.702);

\path [draw=color0, line width=0.04pt]
(axis cs:1,123.629)
--(axis cs:1,163.401);

\path [draw=color0, line width=0.04pt]
(axis cs:2,138.997)
--(axis cs:2,191.651);

\path [draw=color0, line width=0.04pt]
(axis cs:3,152.174)
--(axis cs:3,214.41);

\path [draw=color0, line width=0.04pt]
(axis cs:4,183.426)
--(axis cs:4,237.67);

\path [draw=color0, line width=0.04pt]
(axis cs:5,435.685)
--(axis cs:5,589.329);

\addplot [very thick, color0, mark=-, mark size=3, mark options={solid}, only marks]
table {%
0 130.56
1 123.629
2 138.997
3 152.174
4 183.426
5 435.685
};
\addplot [very thick, color0, mark=-, mark size=3, mark options={solid}, only marks]
table {%
0 152.702
1 163.401
2 191.651
3 214.41
4 237.67
5 589.329
};
\path [draw=color1, line width=0.04pt]
(axis cs:0,53.213)
--(axis cs:0,61.297);

\path [draw=color1, line width=0.04pt]
(axis cs:1,51.051)
--(axis cs:1,58.191);

\path [draw=color1, line width=0.04pt]
(axis cs:2,20.622)
--(axis cs:2,28.386);

\path [draw=color1, line width=0.04pt]
(axis cs:3,16.117)
--(axis cs:3,25.549);

\path [draw=color1, line width=0.04pt]
(axis cs:4,17.108)
--(axis cs:4,23.934);

\path [draw=color1, line width=0.04pt]
(axis cs:5,42.551)
--(axis cs:5,67.985);

\addplot [very thick, color1, mark=-, mark size=3, mark options={solid}, only marks]
table {%
0 53.213
1 51.051
2 20.622
3 16.117
4 17.108
5 42.551
};
\addplot [very thick, color1, mark=-, mark size=3, mark options={solid}, only marks]
table {%
0 61.297
1 58.191
2 28.386
3 25.549
4 23.934
5 67.985
};
\path [draw=color2, line width=0.04pt]
(axis cs:0,51.308)
--(axis cs:0,58.63);

\path [draw=color2, line width=0.04pt]
(axis cs:1,51.22)
--(axis cs:1,58.132);

\path [draw=color2, line width=0.04pt]
(axis cs:2,15.416)
--(axis cs:2,20.408);

\path [draw=color2, line width=0.04pt]
(axis cs:3,9.891)
--(axis cs:3,11.757);

\path [draw=color2, line width=0.04pt]
(axis cs:4,5.793)
--(axis cs:4,10.043);

\path [draw=color2, line width=0.04pt]
(axis cs:5,6.019)
--(axis cs:5,9.693);

\addplot [very thick, color2, mark=-, mark size=3, mark options={solid}, only marks]
table {%
0 51.308
1 51.22
2 15.416
3 9.891
4 5.793
5 6.019
};
\addplot [very thick, color2, mark=-, mark size=3, mark options={solid}, only marks]
table {%
0 58.63
1 58.132
2 20.408
3 11.757
4 10.043
5 9.693
};
\path [draw=color3, line width=0.04pt]
(axis cs:0,51.747)
--(axis cs:0,60.049);

\path [draw=color3, line width=0.04pt]
(axis cs:1,51.715)
--(axis cs:1,55.479);

\path [draw=color3, line width=0.04pt]
(axis cs:2,15.062)
--(axis cs:2,20.354);

\path [draw=color3, line width=0.04pt]
(axis cs:3,8.38)
--(axis cs:3,12.2);

\path [draw=color3, line width=0.04pt]
(axis cs:4,6.151)
--(axis cs:4,9.047);

\path [draw=color3, line width=0.04pt]
(axis cs:5,0.268)
--(axis cs:5,1.436);

\addplot [very thick, color3, mark=-, mark size=3, mark options={solid}, only marks]
table {%
0 51.747
1 51.715
2 15.062
3 8.38
4 6.151
5 0.268
};
\addplot [very thick, color3, mark=-, mark size=3, mark options={solid}, only marks]
table {%
0 60.049
1 55.479
2 20.354
3 12.2
4 9.047
5 1.436
};
\path [draw=color4, line width=0.04pt]
(axis cs:0,51.212)
--(axis cs:0,59.248);

\path [draw=color4, line width=0.04pt]
(axis cs:1,52.368)
--(axis cs:1,57.936);

\path [draw=color4, line width=0.04pt]
(axis cs:2,17.003)
--(axis cs:2,21.603);

\path [draw=color4, line width=0.04pt]
(axis cs:3,10.142)
--(axis cs:3,13.602);

\path [draw=color4, line width=0.04pt]
(axis cs:4,5.592)
--(axis cs:4,8.126);

\path [draw=color4, line width=0.04pt]
(axis cs:5,0)
--(axis cs:5,0);

\addplot [very thick, color4, mark=-, mark size=3, mark options={solid}, only marks]
table {%
0 51.212
1 52.368
2 17.003
3 10.142
4 5.592
5 0
};
\addplot [very thick, color4, mark=-, mark size=3, mark options={solid}, only marks]
table {%
0 59.248
1 57.936
2 21.603
3 13.602
4 8.126
5 0
};
\path [draw=white!27.843137254902!black, line width=0.04pt]
(axis cs:0,50.107)
--(axis cs:0,58.585);

\path [draw=white!27.843137254902!black, line width=0.04pt]
(axis cs:1,52.237)
--(axis cs:1,58.593);

\path [draw=white!27.843137254902!black, line width=0.04pt]
(axis cs:2,15.962)
--(axis cs:2,19.57);

\path [draw=white!27.843137254902!black, line width=0.04pt]
(axis cs:3,9.241)
--(axis cs:3,11.263);

\path [draw=white!27.843137254902!black, line width=0.04pt]
(axis cs:4,6.621)
--(axis cs:4,8.811);

\path [draw=white!27.843137254902!black, line width=0.04pt]
(axis cs:5,0)
--(axis cs:5,0);

\addplot [very thick, white!27.843137254902!black, mark=-, mark size=3, mark options={solid}, only marks]
table {%
0 50.107
1 52.237
2 15.962
3 9.241
4 6.621
5 0
};
\addplot [very thick, white!27.843137254902!black, mark=-, mark size=3, mark options={solid}, only marks]
table {%
0 58.585
1 58.593
2 19.57
3 11.263
4 8.811
5 0
};
\addplot [very thick, color0, dash pattern=on 1pt off 2pt, mark=square*, mark size=0.5, mark options={solid}]
table {%
0 141.631
1 143.515
2 165.324
3 183.292
4 210.548
5 512.507
};
\addplot [very thick, color1, dash pattern=on 3pt off 1pt on 1pt off 1pt on 1pt off 1pt, mark=square*, mark size=0.5, mark options={solid}]
table {%
0 57.255
1 54.621
2 24.504
3 20.833
4 20.521
5 55.268
};
\addplot [very thick, color2, dash pattern=on 5pt off 1pt, mark=square*, mark size=0.5, mark options={solid}]
table {%
0 54.969
1 54.676
2 17.912
3 10.824
4 7.918
5 7.856
};
\addplot [very thick, color3, dash pattern=on 1pt off 3pt on 3pt off 3pt, mark=square*, mark size=0.5, mark options={solid}]
table {%
0 55.898
1 53.597
2 17.708
3 10.29
4 7.599
5 0.852
};
\addplot [very thick, color4, dash pattern=on 1pt off 1pt, mark=square*, mark size=0.5, mark options={solid}]
table {%
0 55.23
1 55.152
2 19.303
3 11.872
4 6.859
5 0
};
\addplot [very thick, white!27.843137254902!black, mark=square*, mark size=0.5, mark options={solid}]
table {%
0 54.346
1 55.415
2 17.766
3 10.252
4 7.716
5 0
};
\end{axis}

\end{tikzpicture}
	\end{minipage}
	\begin{minipage}[t]{\textwidth}
		\vspace{-1em}
		\centering
\begin{tikzpicture}[scale=0.6]
	
	\definecolor{color0}{rgb}{0.0470588235294118,0.364705882352941,0.647058823529412}
	\definecolor{color1}{rgb}{0,0.725490196078431,0.270588235294118}
	\definecolor{color2}{rgb}{1,0.584313725490196,0}
	\definecolor{color3}{rgb}{1,0.172549019607843,0}
	\definecolor{color4}{rgb}{0.717647058823529,0.454901960784314,0.83921568627451}
	
	\begin{axis}[
		hide axis,
		xmin=0,
		xmax=1,
		ymin=0,
		ymax=1,
		legend cell align={left},
		legend columns=7,
		legend style={
			fill opacity=0.8,
			draw opacity=1,
			text opacity=1,
			draw=white!80!black
		}
		]
		\addlegendimage{empty legend}
		\addlegendentry{$\varepsilon$: }
		\addlegendimage{very thick, color0, dash pattern=on 1pt off 2pt}
		\addlegendentry{0.01}
		\addlegendimage{very thick, color1, dash pattern=on 3pt off 1pt on 1pt 
		off 1pt on 1pt off 1pt}
		\addlegendentry{0.1}
		\addlegendimage{very thick, color2, dash pattern=on 5pt off 1pt}
		\addlegendentry{1}
		\addlegendimage{very thick, color3, dash pattern=on 1pt off 3pt on 3pt 	
		off 3pt}
		\addlegendentry{10}
		\addlegendimage{very thick, color4, dash pattern=on 1pt off 1pt}
		\addlegendentry{100}
		\addlegendimage{very thick, white!27.843137254902!black}
		\addlegendentry{withoutDp}
		\addlegendimage{very thick, color0, dash pattern=on 1pt off 2pt, 
		mark=square*, mark size=0.5, mark options={solid}, forget plot}
		
	\end{axis}
	
\end{tikzpicture}
	\end{minipage}
	\begin{minipage}[t]{\textwidth}
		\captionsetup[sub]{labelformat=parens}
		\begingroup
		\captionsetup{type=figure}
		\begin{subfigure}{0.3\textwidth}
			\vspace*{-5cm}
			\caption{\tt GEOLIFE}
		\end{subfigure}
		\begin{subfigure}{0.3\textwidth}
			\vspace*{-5cm}
			\caption{\tt MADRID}
		\end{subfigure}
		\begin{subfigure}{0.3\textwidth}
			\vspace*{-5cm}
			\caption{\tt BERLIN}
		\end{subfigure}
		\endgroup
	\end{minipage}
	\begin{minipage}[t]{\textwidth}
		\vspace*{-2.7cm}
		\caption{ $\mathrm{LocationError}$ for different $M$ and $\varepsilon$
			for each data set. The standard deviation of all 10 runs is 
			represented by the error bars.}
		\label{fig:emd_tile_counts}
	\end{minipage}
\vspace*{-2.5cm}
\end{figure}

In
Figure~\ref{fig:emd_tile_counts} we
show the impact of $M$ and $\varepsilon$ on the $\mathrm{LocationError}$.
Note that the range of the $y$-axis 
differs greatly between data sets since the $\mathrm{LocationError}$ depends on the 
underlying extent and split of the tessellation and the uniformity of the geospatial distribution. Recall 
that the $\mathrm{LocationError}$ can be 
interpreted as the distance every visit needs to be moved on average 
to create the noisy distribution.
The error for \textit{withoutDp} is remarkably 
low for all three data sets, even for $M=1$. For {\tt MADRID}, the error is 
only 200 meters and 
for {\tt BERLIN} $\approx$ 60 meters. For {\tt GEOLIFE}, the error is 2,290 
meters which is rather high 
compared to the other two data sets. But considering that a sampling of $M = 1$ 
only contains 1\,\% (182 
trips) of the {\tt GEOLIFE} data set, one could argue that this error, which 
corresponds roughly 
to the diameter of one tile, still lies within a reasonable range. 
This indicates that there is not a 
large information gain when increasing $M$ to capture the geospatial distribution, confirming
insights from mobility research stated in Section~\ref{sec:boundedContr} 
that people mostly visit a few recurrent 
locations. Overall patterns can therefore already be captured with only a fraction of the actual trips.
For all three data sets, the maximum $M$ highly increases the error for most 
variations of $\varepsilon$,
indicating the usefulness of cut off values.
E.g., for {\tt BERLIN} the curves corresponding to $\varepsilon \geq 1$ follow the course of 
\textit{withoutDp}, but for 
stronger privacy guarantees, such as $\varepsilon = 0.1$, we see a tipping 
point of the error at $M = 6$. 
Respectively, for {\tt MADRID} there is a tipping point for $\varepsilon = 1$ 
and $M = 5$. 
Again, we see the strongest effect of the sensitivity onto the error for
{\tt GEOLIFE}. It is 
striking that even for a high privacy budget of $\varepsilon = 100$ the error deviates from 
\textit{withoutDp} at $M = 2,153$. 
For $\varepsilon = 10$ the error increases for any $M > 1$, while for all $\varepsilon < 10$ the error for 
$M = 1$ is already a multitude higher than the \textit{withoutDp} baseline, only increasing or staying 
constant for higher $M$-values. This suggests that these 
$\varepsilon$-values are not suitable for meaningful analyses of \textit{visits 
per location} for {\tt 
GEOLIFE}.

\begin{figure}[tb]
	\centering
	\begin{minipage}[t]{0.3\textwidth}	
\begin{tikzpicture}[scale=0.6]

\definecolor{color0}{rgb}{0.0470588235294118,0.364705882352941,0.647058823529412}
\definecolor{color1}{rgb}{0,0.725490196078431,0.270588235294118}
\definecolor{color2}{rgb}{1,0.584313725490196,0}
\definecolor{color3}{rgb}{1,0.172549019607843,0}
\definecolor{color4}{rgb}{0.717647058823529,0.454901960784314,0.83921568627451}

\begin{axis}[
legend cell align={left},
legend columns=6,
legend style={
  fill opacity=0.8,
  draw opacity=1,
  text opacity=1,
  at={(1.1,-0.15)},
  anchor=south east,
  draw=white!80!black
},
scaled y ticks=manual:{}{\pgfmathparse{#1}},
tick pos=both,
x grid style={white!69.0196078431373!black},
xlabel={M},
xmin=-0.3, xmax=6.3,
xtick style={color=black},
xtick={0,1,2,3,4,5,6},
xtick={0,1,2,3,4,5,6},
xtick={0,1,2,3,4,5,6},
xtick={0,1,2,3,4,5,6},
xtick={0,1,2,3,4,5,6},
xtick={0,1,2,3,4,5,6},
xtick={0,1,2,3,4,5,6},
xtick={0,1,2,3,4,5,6},
xtick={0,1,2,3,4,5,6},
xtick={0,1,2,3,4,5,6},
xtick={0,1,2,3,4,5,6},
xtick={0,1,2,3,4,5,6},
xtick={0,1,2,3,4,5,6},
xtick={0,1,2,3,4,5,6},
xtick={0,1,2,3,4,5,6},
xtick={0,1,2,3,4,5,6},
xtick={0,1,2,3,4,5,6},
xtick={0,1,2,3,4,5,6},
xtick={0,1,2,3,4,5,6},
xtick={0,1,2,3,4,5,6},
xtick={0,1,2,3,4,5,6},
xtick={0,1,2,3,4,5,6},
xtick={0,1,2,3,4,5,6},
xtick={0,1,2,3,4,5,6},
xticklabels={1,3,8,28,90,215,2153},
xticklabels={1,3,8,28,90,215,2153},
xticklabels={1,3,8,28,90,215,2153},
xticklabels={1,3,8,28,90,215,2153},
xticklabels={1,3,8,28,90,215,2153},
xticklabels={1,3,8,28,90,215,2153},
xticklabels={1,3,8,28,90,215,2153},
xticklabels={1,3,8,28,90,215,2153},
xticklabels={1,3,8,28,90,215,2153},
xticklabels={1,3,8,28,90,215,2153},
xticklabels={1,3,8,28,90,215,2153},
xticklabels={1,3,8,28,90,215,2153},
xticklabels={1,3,8,28,90,215,2153},
xticklabels={1,3,8,28,90,215,2153},
xticklabels={1,3,8,28,90,215,2153},
xticklabels={1,3,8,28,90,215,2153},
xticklabels={1,3,8,28,90,215,2153},
xticklabels={1,3,8,28,90,215,2153},
xticklabels={1,3,8,28,90,215,2153},
xticklabels={1,3,8,28,90,215,2153},
xticklabels={1,3,8,28,90,215,2153},
xticklabels={1,3,8,28,90,215,2153},
xticklabels={1,3,8,28,90,215,2153},
xticklabels={1,3,8,28,90,215,2153},
y grid style={white!69.0196078431373!black},
ylabel={OdFlowError},
ymin=0, ymax=2.2,
ytick style={color=black},
]
\path [draw=color0, line width=0.04pt]
(axis cs:0,2)
--(axis cs:0,2);

\path [draw=color0, line width=0.04pt]
(axis cs:1,2)
--(axis cs:1,2);

\path [draw=color0, line width=0.04pt]
(axis cs:2,2)
--(axis cs:2,2);

\path [draw=color0, line width=0.04pt]
(axis cs:3,2)
--(axis cs:3,2);

\path [draw=color0, line width=0.04pt]
(axis cs:4,2)
--(axis cs:4,2);

\path [draw=color0, line width=0.04pt]
(axis cs:5,2)
--(axis cs:5,2);

\path [draw=color0, line width=0.04pt]
(axis cs:6,2)
--(axis cs:6,2);

\addplot [very thick, color0, mark=-, mark size=3, mark options={solid}, only marks, forget plot]
table {%
0 2
1 2
2 2
3 2
4 2
5 2
6 2
};
\addplot [very thick, color0, mark=-, mark size=3, mark options={solid}, only marks, forget plot]
table {%
0 2
1 2
2 2
3 2
4 2
5 2
6 2
};
\path [draw=color1, line width=0.04pt]
(axis cs:0,2)
--(axis cs:0,2);

\path [draw=color1, line width=0.04pt]
(axis cs:1,2)
--(axis cs:1,2);

\path [draw=color1, line width=0.04pt]
(axis cs:2,2)
--(axis cs:2,2);

\path [draw=color1, line width=0.04pt]
(axis cs:3,2)
--(axis cs:3,2);

\path [draw=color1, line width=0.04pt]
(axis cs:4,2)
--(axis cs:4,2);

\path [draw=color1, line width=0.04pt]
(axis cs:5,2)
--(axis cs:5,2);

\path [draw=color1, line width=0.04pt]
(axis cs:6,2)
--(axis cs:6,2);

\addplot [very thick, color1, mark=-, mark size=3, mark options={solid}, only marks, forget plot]
table {%
0 2
1 2
2 2
3 2
4 2
5 2
6 2
};
\addplot [very thick, color1, mark=-, mark size=3, mark options={solid}, only marks, forget plot]
table {%
0 2
1 2
2 2
3 2
4 2
5 2
6 2
};
\path [draw=color2, line width=0.04pt]
(axis cs:0,2)
--(axis cs:0,2);

\path [draw=color2, line width=0.04pt]
(axis cs:1,2)
--(axis cs:1,2);

\path [draw=color2, line width=0.04pt]
(axis cs:2,2)
--(axis cs:2,2);

\path [draw=color2, line width=0.04pt]
(axis cs:3,2)
--(axis cs:3,2);

\path [draw=color2, line width=0.04pt]
(axis cs:4,2)
--(axis cs:4,2);

\path [draw=color2, line width=0.04pt]
(axis cs:5,2)
--(axis cs:5,2);

\path [draw=color2, line width=0.04pt]
(axis cs:6,2)
--(axis cs:6,2);

\addplot [very thick, color2, mark=-, mark size=3, mark options={solid}, only marks, forget plot]
table {%
0 2
1 2
2 2
3 2
4 2
5 2
6 2
};
\addplot [very thick, color2, mark=-, mark size=3, mark options={solid}, only marks, forget plot]
table {%
0 2
1 2
2 2
3 2
4 2
5 2
6 2
};
\path [draw=color3, line width=0.04pt]
(axis cs:0,1.983)
--(axis cs:0,1.985);

\path [draw=color3, line width=0.04pt]
(axis cs:1,1.998)
--(axis cs:1,1.998);

\path [draw=color3, line width=0.04pt]
(axis cs:2,1.999)
--(axis cs:2,1.999);

\path [draw=color3, line width=0.04pt]
(axis cs:3,2)
--(axis cs:3,2);

\path [draw=color3, line width=0.04pt]
(axis cs:4,2)
--(axis cs:4,2);

\path [draw=color3, line width=0.04pt]
(axis cs:5,2)
--(axis cs:5,2);

\path [draw=color3, line width=0.04pt]
(axis cs:6,2)
--(axis cs:6,2);

\addplot [very thick, color3, mark=-, mark size=3, mark options={solid}, only marks, forget plot]
table {%
0 1.983
1 1.998
2 1.999
3 2
4 2
5 2
6 2
};
\addplot [very thick, color3, mark=-, mark size=3, mark options={solid}, only marks, forget plot]
table {%
0 1.985
1 1.998
2 1.999
3 2
4 2
5 2
6 2
};
\path [draw=color4, line width=0.04pt]
(axis cs:0,1.973)
--(axis cs:0,1.977);

\path [draw=color4, line width=0.04pt]
(axis cs:1,1.932)
--(axis cs:1,1.936);

\path [draw=color4, line width=0.04pt]
(axis cs:2,1.868)
--(axis cs:2,1.876);

\path [draw=color4, line width=0.04pt]
(axis cs:3,1.995)
--(axis cs:3,1.995);

\path [draw=color4, line width=0.04pt]
(axis cs:4,1.999)
--(axis cs:4,1.999);

\path [draw=color4, line width=0.04pt]
(axis cs:5,1.999)
--(axis cs:5,1.999);

\path [draw=color4, line width=0.04pt]
(axis cs:6,2)
--(axis cs:6,2);

\addplot [very thick, color4, mark=-, mark size=3, mark options={solid}, only marks, forget plot]
table {%
0 1.973
1 1.932
2 1.868
3 1.995
4 1.999
5 1.999
6 2
};
\addplot [very thick, color4, mark=-, mark size=3, mark options={solid}, only marks, forget plot]
table {%
0 1.977
1 1.936
2 1.876
3 1.995
4 1.999
5 1.999
6 2
};
\path [draw=white!27.843137254902!black, line width=0.04pt]
(axis cs:0,1.974)
--(axis cs:0,1.976);

\path [draw=white!27.843137254902!black, line width=0.04pt]
(axis cs:1,1.928)
--(axis cs:1,1.934);

\path [draw=white!27.843137254902!black, line width=0.04pt]
(axis cs:2,1.853)
--(axis cs:2,1.861);

\path [draw=white!27.843137254902!black, line width=0.04pt]
(axis cs:3,1.661)
--(axis cs:3,1.677);

\path [draw=white!27.843137254902!black, line width=0.04pt]
(axis cs:4,1.292)
--(axis cs:4,1.306);

\path [draw=white!27.843137254902!black, line width=0.04pt]
(axis cs:5,0.912)
--(axis cs:5,0.928);

\path [draw=white!27.843137254902!black, line width=0.04pt]
(axis cs:6,0)
--(axis cs:6,0);

\addplot [very thick, white!27.843137254902!black, mark=-, mark size=3, mark options={solid}, only marks, forget plot]
table {%
0 1.974
1 1.928
2 1.853
3 1.661
4 1.292
5 0.912
6 0
};
\addplot [very thick, white!27.843137254902!black, mark=-, mark size=3, mark options={solid}, only marks, forget plot]
table {%
0 1.976
1 1.934
2 1.861
3 1.677
4 1.306
5 0.928
6 0
};
\addplot [very thick, color0, dash pattern=on 1pt off 2pt]
table {%
0 2
1 2
2 2
3 2
4 2
5 2
6 2
};

\addplot [very thick, color1, dash pattern=on 3pt off 1pt on 1pt off 1pt on 1pt off 1pt]
table {%
0 2
1 2
2 2
3 2
4 2
5 2
6 2
};

\addplot [very thick, color2, dash pattern=on 5pt off 1pt]
table {%
0 2
1 2
2 2
3 2
4 2
5 2
6 2
};

\addplot [very thick, color3, dash pattern=on 1pt off 3pt on 3pt off 3pt]
table {%
0 1.984
1 1.998
2 1.999
3 2
4 2
5 2
6 2
};

\addplot [very thick, color4, dash pattern=on 1pt off 1pt]
table {%
0 1.975
1 1.934
2 1.872
3 1.995
4 1.999
5 1.999
6 2
};

\addplot [very thick, white!27.843137254902!black]
table {%
0 1.975
1 1.931
2 1.857
3 1.669
4 1.299
5 0.92
6 0
};

\addplot [very thick, color0, dash pattern=on 1pt off 2pt, mark=square*, mark size=0.5, mark options={solid}, forget plot]
table {%
0 2
1 2
2 2
3 2
4 2
5 2
6 2
};
\addplot [very thick, color1, dash pattern=on 3pt off 1pt on 1pt off 1pt on 1pt off 1pt, mark=square*, mark size=0.5, mark options={solid}, forget plot]
table {%
0 2
1 2
2 2
3 2
4 2
5 2
6 2
};
\addplot [very thick, color2, dash pattern=on 5pt off 1pt, mark=square*, mark size=0.5, mark options={solid}, forget plot]
table {%
0 2
1 2
2 2
3 2
4 2
5 2
6 2
};
\addplot [very thick, color3, dash pattern=on 1pt off 3pt on 3pt off 3pt, mark=square*, mark size=0.5, mark options={solid}, forget plot]
table {%
0 1.984
1 1.998
2 1.999
3 2
4 2
5 2
6 2
};
\addplot [very thick, color4, dash pattern=on 1pt off 1pt, mark=square*, mark size=0.5, mark options={solid}, forget plot]
table {%
0 1.975
1 1.934
2 1.872
3 1.995
4 1.999
5 1.999
6 2
};
\addplot [very thick, white!27.843137254902!black, mark=square*, mark size=0.5, mark options={solid}, forget plot]
table {%
0 1.975
1 1.931
2 1.857
3 1.669
4 1.299
5 0.92
6 0
};
\end{axis}

\end{tikzpicture}
	\end{minipage}
	\hspace{2em}
	\begin{minipage}[t]{0.3\textwidth}
\begin{tikzpicture}[scale=0.6]

\definecolor{color0}{rgb}{0.0470588235294118,0.364705882352941,0.647058823529412}
\definecolor{color1}{rgb}{0,0.725490196078431,0.270588235294118}
\definecolor{color2}{rgb}{1,0.584313725490196,0}
\definecolor{color3}{rgb}{1,0.172549019607843,0}
\definecolor{color4}{rgb}{0.717647058823529,0.454901960784314,0.83921568627451}

\begin{axis}[
tick pos=both,
x grid style={white!69.0196078431373!black},
xlabel={M},
xmin=-0.2, xmax=4.2,
xtick style={color=black},
xtick={0,1,2,3,4},
xtick={0,1,2,3,4},
xtick={0,1,2,3,4},
xtick={0,1,2,3,4},
xtick={0,1,2,3,4},
xtick={0,1,2,3,4},
xtick={0,1,2,3,4},
xtick={0,1,2,3,4},
xtick={0,1,2,3,4},
xtick={0,1,2,3,4},
xtick={0,1,2,3,4},
xtick={0,1,2,3,4},
xtick={0,1,2,3,4},
xtick={0,1,2,3,4},
xtick={0,1,2,3,4},
xtick={0,1,2,3,4},
xtick={0,1,2,3,4},
xtick={0,1,2,3,4},
xticklabels={1,2,4,5,20},
xticklabels={1,2,4,5,20},
xticklabels={1,2,4,5,20},
xticklabels={1,2,4,5,20},
xticklabels={1,2,4,5,20},
xticklabels={1,2,4,5,20},
xticklabels={1,2,4,5,20},
xticklabels={1,2,4,5,20},
xticklabels={1,2,4,5,20},
xticklabels={1,2,4,5,20},
xticklabels={1,2,4,5,20},
xticklabels={1,2,4,5,20},
xticklabels={1,2,4,5,20},
xticklabels={1,2,4,5,20},
xticklabels={1,2,4,5,20},
xticklabels={1,2,4,5,20},
xticklabels={1,2,4,5,20},
xticklabels={1,2,4,5,20},
y grid style={white!69.0196078431373!black},
ymin=0, ymax=2.2,
ytick style={color=black}
]
\path [draw=color0, line width=0.04pt]
(axis cs:0,1.965)
--(axis cs:0,1.965);

\path [draw=color0, line width=0.04pt]
(axis cs:1,1.965)
--(axis cs:1,1.965);

\path [draw=color0, line width=0.04pt]
(axis cs:2,1.965)
--(axis cs:2,1.967);

\path [draw=color0, line width=0.04pt]
(axis cs:3,1.965)
--(axis cs:3,1.965);

\path [draw=color0, line width=0.04pt]
(axis cs:4,1.966)
--(axis cs:4,1.966);

\addplot [very thick, color0, mark=-, mark size=3, mark options={solid}, only marks]
table {%
0 1.965
1 1.965
2 1.965
3 1.965
4 1.966
};
\addplot [very thick, color0, mark=-, mark size=3, mark options={solid}, only marks]
table {%
0 1.965
1 1.965
2 1.967
3 1.965
4 1.966
};
\path [draw=color1, line width=0.04pt]
(axis cs:0,1.962)
--(axis cs:0,1.962);

\path [draw=color1, line width=0.04pt]
(axis cs:1,1.963)
--(axis cs:1,1.963);

\path [draw=color1, line width=0.04pt]
(axis cs:2,1.964)
--(axis cs:2,1.964);

\path [draw=color1, line width=0.04pt]
(axis cs:3,1.964)
--(axis cs:3,1.964);

\path [draw=color1, line width=0.04pt]
(axis cs:4,1.965)
--(axis cs:4,1.965);

\addplot [very thick, color1, mark=-, mark size=3, mark options={solid}, only marks]
table {%
0 1.962
1 1.963
2 1.964
3 1.964
4 1.965
};
\addplot [very thick, color1, mark=-, mark size=3, mark options={solid}, only marks]
table {%
0 1.962
1 1.963
2 1.964
3 1.964
4 1.965
};
\path [draw=color2, line width=0.04pt]
(axis cs:0,1.92)
--(axis cs:0,1.92);

\path [draw=color2, line width=0.04pt]
(axis cs:1,1.936)
--(axis cs:1,1.936);

\path [draw=color2, line width=0.04pt]
(axis cs:2,1.949)
--(axis cs:2,1.949);

\path [draw=color2, line width=0.04pt]
(axis cs:3,1.953)
--(axis cs:3,1.953);

\path [draw=color2, line width=0.04pt]
(axis cs:4,1.963)
--(axis cs:4,1.963);

\addplot [very thick, color2, mark=-, mark size=3, mark options={solid}, only marks]
table {%
0 1.92
1 1.936
2 1.949
3 1.953
4 1.963
};
\addplot [very thick, color2, mark=-, mark size=3, mark options={solid}, only marks]
table {%
0 1.92
1 1.936
2 1.949
3 1.953
4 1.963
};
\path [draw=color3, line width=0.04pt]
(axis cs:0,1.411)
--(axis cs:0,1.413);

\path [draw=color3, line width=0.04pt]
(axis cs:1,1.144)
--(axis cs:1,1.146);

\path [draw=color3, line width=0.04pt]
(axis cs:2,1.633)
--(axis cs:2,1.635);

\path [draw=color3, line width=0.04pt]
(axis cs:3,1.726)
--(axis cs:3,1.728);

\path [draw=color3, line width=0.04pt]
(axis cs:4,1.927)
--(axis cs:4,1.927);

\addplot [very thick, color3, mark=-, mark size=3, mark options={solid}, only marks]
table {%
0 1.411
1 1.144
2 1.633
3 1.726
4 1.927
};
\addplot [very thick, color3, mark=-, mark size=3, mark options={solid}, only marks]
table {%
0 1.413
1 1.146
2 1.635
3 1.728
4 1.927
};
\path [draw=color4, line width=0.04pt]
(axis cs:0,1.397)
--(axis cs:0,1.399);

\path [draw=color4, line width=0.04pt]
(axis cs:1,0.7)
--(axis cs:1,0.702);

\path [draw=color4, line width=0.04pt]
(axis cs:2,0.171)
--(axis cs:2,0.173);

\path [draw=color4, line width=0.04pt]
(axis cs:3,0.09)
--(axis cs:3,0.092);

\path [draw=color4, line width=0.04pt]
(axis cs:4,0.97)
--(axis cs:4,0.976);

\addplot [very thick, color4, mark=-, mark size=3, mark options={solid}, only marks]
table {%
0 1.397
1 0.7
2 0.171
3 0.09
4 0.97
};
\addplot [very thick, color4, mark=-, mark size=3, mark options={solid}, only marks]
table {%
0 1.399
1 0.702
2 0.173
3 0.092
4 0.976
};
\path [draw=white!27.843137254902!black, line width=0.04pt]
(axis cs:0,1.396)
--(axis cs:0,1.398);

\path [draw=white!27.843137254902!black, line width=0.04pt]
(axis cs:1,0.701)
--(axis cs:1,0.703);

\path [draw=white!27.843137254902!black, line width=0.04pt]
(axis cs:2,0.172)
--(axis cs:2,0.174);

\path [draw=white!27.843137254902!black, line width=0.04pt]
(axis cs:3,0.089)
--(axis cs:3,0.091);

\path [draw=white!27.843137254902!black, line width=0.04pt]
(axis cs:4,0)
--(axis cs:4,0);

\addplot [very thick, white!27.843137254902!black, mark=-, mark size=3, mark options={solid}, only marks]
table {%
0 1.396
1 0.701
2 0.172
3 0.089
4 0
};
\addplot [very thick, white!27.843137254902!black, mark=-, mark size=3, mark options={solid}, only marks]
table {%
0 1.398
1 0.703
2 0.174
3 0.091
4 0
};
\addplot [very thick, color0, dash pattern=on 1pt off 2pt, mark=square*, mark size=0.5, mark options={solid}]
table {%
0 1.965
1 1.965
2 1.966
3 1.965
4 1.966
};
\addplot [very thick, color1, dash pattern=on 3pt off 1pt on 1pt off 1pt on 1pt off 1pt, mark=square*, mark size=0.5, mark options={solid}]
table {%
0 1.962
1 1.963
2 1.964
3 1.964
4 1.965
};
\addplot [very thick, color2, dash pattern=on 5pt off 1pt, mark=square*, mark size=0.5, mark options={solid}]
table {%
0 1.92
1 1.936
2 1.949
3 1.953
4 1.963
};
\addplot [very thick, color3, dash pattern=on 1pt off 3pt on 3pt off 3pt, mark=square*, mark size=0.5, mark options={solid}]
table {%
0 1.412
1 1.145
2 1.634
3 1.727
4 1.927
};
\addplot [very thick, color4, dash pattern=on 1pt off 1pt, mark=square*, mark size=0.5, mark options={solid}]
table {%
0 1.398
1 0.701
2 0.172
3 0.091
4 0.973
};
\addplot [very thick, white!27.843137254902!black, mark=square*, mark size=0.5, mark options={solid}]
table {%
0 1.397
1 0.702
2 0.173
3 0.09
4 0
};
\end{axis}

\end{tikzpicture}
	\end{minipage}
	\hspace{1em}
	\begin{minipage}[t]{0.3\textwidth}
\begin{tikzpicture}[scale=0.6]

\definecolor{color0}{rgb}{0.0470588235294118,0.364705882352941,0.647058823529412}
\definecolor{color1}{rgb}{0,0.725490196078431,0.270588235294118}
\definecolor{color2}{rgb}{1,0.584313725490196,0}
\definecolor{color3}{rgb}{1,0.172549019607843,0}
\definecolor{color4}{rgb}{0.717647058823529,0.454901960784314,0.83921568627451}

\begin{axis}[
tick pos=both,
x grid style={white!69.0196078431373!black},
xlabel={M},
xmin=-0.25, xmax=5.25,
xtick style={color=black},
xtick={0,1,2,3,4,5},
xtick={0,1,2,3,4,5},
xtick={0,1,2,3,4,5},
xtick={0,1,2,3,4,5},
xtick={0,1,2,3,4,5},
xtick={0,1,2,3,4,5},
xtick={0,1,2,3,4,5},
xtick={0,1,2,3,4,5},
xtick={0,1,2,3,4,5},
xtick={0,1,2,3,4,5},
xtick={0,1,2,3,4,5},
xtick={0,1,2,3,4,5},
xtick={0,1,2,3,4,5},
xtick={0,1,2,3,4,5},
xtick={0,1,2,3,4,5},
xtick={0,1,2,3,4,5},
xtick={0,1,2,3,4,5},
xtick={0,1,2,3,4,5},
xticklabels={1,2,4,5,6,16},
xticklabels={1,2,4,5,6,16},
xticklabels={1,2,4,5,6,16},
xticklabels={1,2,4,5,6,16},
xticklabels={1,2,4,5,6,16},
xticklabels={1,2,4,5,6,16},
xticklabels={1,2,4,5,6,16},
xticklabels={1,2,4,5,6,16},
xticklabels={1,2,4,5,6,16},
xticklabels={1,2,4,5,6,16},
xticklabels={1,2,4,5,6,16},
xticklabels={1,2,4,5,6,16},
xticklabels={1,2,4,5,6,16},
xticklabels={1,2,4,5,6,16},
xticklabels={1,2,4,5,6,16},
xticklabels={1,2,4,5,6,16},
xticklabels={1,2,4,5,6,16},
xticklabels={1,2,4,5,6,16},
y grid style={white!69.0196078431373!black},
ymin=0, ymax=2.2,
ytick style={color=black}
]
\path [draw=color0, line width=0.04pt]
(axis cs:0,1.653)
--(axis cs:0,1.655);

\path [draw=color0, line width=0.04pt]
(axis cs:1,1.654)
--(axis cs:1,1.656);

\path [draw=color0, line width=0.04pt]
(axis cs:2,1.656)
--(axis cs:2,1.66);

\path [draw=color0, line width=0.04pt]
(axis cs:3,1.658)
--(axis cs:3,1.66);

\path [draw=color0, line width=0.04pt]
(axis cs:4,1.658)
--(axis cs:4,1.662);

\path [draw=color0, line width=0.04pt]
(axis cs:5,1.663)
--(axis cs:5,1.667);

\addplot [very thick, color0, mark=-, mark size=3, mark options={solid}, only marks]
table {%
0 1.653
1 1.654
2 1.656
3 1.658
4 1.658
5 1.663
};
\addplot [very thick, color0, mark=-, mark size=3, mark options={solid}, only marks]
table {%
0 1.655
1 1.656
2 1.66
3 1.66
4 1.662
5 1.667
};
\path [draw=color1, line width=0.04pt]
(axis cs:0,1.533)
--(axis cs:0,1.537);

\path [draw=color1, line width=0.04pt]
(axis cs:1,1.53)
--(axis cs:1,1.532);

\path [draw=color1, line width=0.04pt]
(axis cs:2,1.556)
--(axis cs:2,1.56);

\path [draw=color1, line width=0.04pt]
(axis cs:3,1.57)
--(axis cs:3,1.574);

\path [draw=color1, line width=0.04pt]
(axis cs:4,1.583)
--(axis cs:4,1.585);

\path [draw=color1, line width=0.04pt]
(axis cs:5,1.634)
--(axis cs:5,1.638);

\addplot [very thick, color1, mark=-, mark size=3, mark options={solid}, only marks]
table {%
0 1.533
1 1.53
2 1.556
3 1.57
4 1.583
5 1.634
};
\addplot [very thick, color1, mark=-, mark size=3, mark options={solid}, only marks]
table {%
0 1.537
1 1.532
2 1.56
3 1.574
4 1.585
5 1.638
};
\path [draw=color2, line width=0.04pt]
(axis cs:0,1.157)
--(axis cs:0,1.161);

\path [draw=color2, line width=0.04pt]
(axis cs:1,1.05)
--(axis cs:1,1.054);

\path [draw=color2, line width=0.04pt]
(axis cs:2,1.084)
--(axis cs:2,1.086);

\path [draw=color2, line width=0.04pt]
(axis cs:3,1.125)
--(axis cs:3,1.129);

\path [draw=color2, line width=0.04pt]
(axis cs:4,1.164)
--(axis cs:4,1.168);

\path [draw=color2, line width=0.04pt]
(axis cs:5,1.392)
--(axis cs:5,1.396);

\addplot [very thick, color2, mark=-, mark size=3, mark options={solid}, only marks]
table {%
0 1.157
1 1.05
2 1.084
3 1.125
4 1.164
5 1.392
};
\addplot [very thick, color2, mark=-, mark size=3, mark options={solid}, only marks]
table {%
0 1.161
1 1.054
2 1.086
3 1.129
4 1.168
5 1.396
};
\path [draw=color3, line width=0.04pt]
(axis cs:0,0.94)
--(axis cs:0,0.944);

\path [draw=color3, line width=0.04pt]
(axis cs:1,0.569)
--(axis cs:1,0.571);

\path [draw=color3, line width=0.04pt]
(axis cs:2,0.411)
--(axis cs:2,0.415);

\path [draw=color3, line width=0.04pt]
(axis cs:3,0.419)
--(axis cs:3,0.421);

\path [draw=color3, line width=0.04pt]
(axis cs:4,0.44)
--(axis cs:4,0.442);

\path [draw=color3, line width=0.04pt]
(axis cs:5,0.75)
--(axis cs:5,0.752);

\addplot [very thick, color3, mark=-, mark size=3, mark options={solid}, only marks]
table {%
0 0.94
1 0.569
2 0.411
3 0.419
4 0.44
5 0.75
};
\addplot [very thick, color3, mark=-, mark size=3, mark options={solid}, only marks]
table {%
0 0.944
1 0.571
2 0.415
3 0.421
4 0.442
5 0.752
};
\path [draw=color4, line width=0.04pt]
(axis cs:0,0.938)
--(axis cs:0,0.942);

\path [draw=color4, line width=0.04pt]
(axis cs:1,0.523)
--(axis cs:1,0.525);

\path [draw=color4, line width=0.04pt]
(axis cs:2,0.197)
--(axis cs:2,0.197);

\path [draw=color4, line width=0.04pt]
(axis cs:3,0.117)
--(axis cs:3,0.119);

\path [draw=color4, line width=0.04pt]
(axis cs:4,0.061)
--(axis cs:4,0.061);

\path [draw=color4, line width=0.04pt]
(axis cs:5,0.043)
--(axis cs:5,0.045);

\addplot [very thick, color4, mark=-, mark size=3, mark options={solid}, only marks]
table {%
0 0.938
1 0.523
2 0.197
3 0.117
4 0.061
5 0.043
};
\addplot [very thick, color4, mark=-, mark size=3, mark options={solid}, only marks]
table {%
0 0.942
1 0.525
2 0.197
3 0.119
4 0.061
5 0.045
};
\path [draw=white!27.843137254902!black, line width=0.04pt]
(axis cs:0,0.94)
--(axis cs:0,0.942);

\path [draw=white!27.843137254902!black, line width=0.04pt]
(axis cs:1,0.523)
--(axis cs:1,0.525);

\path [draw=white!27.843137254902!black, line width=0.04pt]
(axis cs:2,0.196)
--(axis cs:2,0.198);

\path [draw=white!27.843137254902!black, line width=0.04pt]
(axis cs:3,0.118)
--(axis cs:3,0.118);

\path [draw=white!27.843137254902!black, line width=0.04pt]
(axis cs:4,0.06)
--(axis cs:4,0.062);

\path [draw=white!27.843137254902!black, line width=0.04pt]
(axis cs:5,0)
--(axis cs:5,0);

\addplot [very thick, white!27.843137254902!black, mark=-, mark size=3, mark options={solid}, only marks]
table {%
0 0.94
1 0.523
2 0.196
3 0.118
4 0.06
5 0
};
\addplot [very thick, white!27.843137254902!black, mark=-, mark size=3, mark options={solid}, only marks]
table {%
0 0.942
1 0.525
2 0.198
3 0.118
4 0.062
5 0
};
\addplot [very thick, color0, dash pattern=on 1pt off 2pt, mark=square*, mark size=0.5, mark options={solid}]
table {%
0 1.654
1 1.655
2 1.658
3 1.659
4 1.66
5 1.665
};
\addplot [very thick, color1, dash pattern=on 3pt off 1pt on 1pt off 1pt on 1pt off 1pt, mark=square*, mark size=0.5, mark options={solid}]
table {%
0 1.535
1 1.531
2 1.558
3 1.572
4 1.584
5 1.636
};
\addplot [very thick, color2, dash pattern=on 5pt off 1pt, mark=square*, mark size=0.5, mark options={solid}]
table {%
0 1.159
1 1.052
2 1.085
3 1.127
4 1.166
5 1.394
};
\addplot [very thick, color3, dash pattern=on 1pt off 3pt on 3pt off 3pt, mark=square*, mark size=0.5, mark options={solid}]
table {%
0 0.942
1 0.57
2 0.413
3 0.42
4 0.441
5 0.751
};
\addplot [very thick, color4, dash pattern=on 1pt off 1pt, mark=square*, mark size=0.5, mark options={solid}]
table {%
0 0.94
1 0.524
2 0.197
3 0.118
4 0.061
5 0.044
};
\addplot [very thick, white!27.843137254902!black, mark=square*, mark size=0.5, mark options={solid}]
table {%
0 0.941
1 0.524
2 0.197
3 0.118
4 0.061
5 0
};
\end{axis}

\end{tikzpicture}
	\end{minipage}
	\begin{minipage}[t]{\textwidth}
		\vspace{-1em}
		\centering
\begin{tikzpicture}[scale=0.6]
	
	\definecolor{color0}{rgb}{0.0470588235294118,0.364705882352941,0.647058823529412}
	\definecolor{color1}{rgb}{0,0.725490196078431,0.270588235294118}
	\definecolor{color2}{rgb}{1,0.584313725490196,0}
	\definecolor{color3}{rgb}{1,0.172549019607843,0}
	\definecolor{color4}{rgb}{0.717647058823529,0.454901960784314,0.83921568627451}
	
	\begin{axis}[
		hide axis,
		xmin=0,
		xmax=1,
		ymin=0,
		ymax=1,
		legend cell align={left},
		legend columns=7,
		legend style={
			fill opacity=0.8,
			draw opacity=1,
			text opacity=1,
			draw=white!80!black
		}
		]

		\addlegendimage{empty legend}
		\addlegendentry{$\varepsilon$: }
		
		\addlegendimage{very thick, color0, dash pattern=on 1pt off 2pt}
		\addlegendentry{0.01}
		\addlegendimage{very thick, color1, dash pattern=on 3pt off 1pt on 1pt 
		off 1pt on 1pt off 1pt}
		\addlegendentry{0.1}
		\addlegendimage{very thick, color2, dash pattern=on 5pt off 1pt}
		\addlegendentry{1}
		\addlegendimage{very thick, color3, dash pattern=on 1pt off 3pt on 3pt 
		off 3pt}
		\addlegendentry{10}
		\addlegendimage{very thick, color4, dash pattern=on 1pt off 1pt}
		\addlegendentry{100}
		\addlegendimage{very thick, white!27.843137254902!black}
		\addlegendentry{withoutDp}
		\addlegendimage{very thick, color0, dash pattern=on 1pt off 2pt, 
	mark=square*, mark size=0.5, mark options={solid}, forget plot}

	\end{axis}
	
\end{tikzpicture}
	\end{minipage}
	\begin{minipage}[t]{\textwidth}
		\captionsetup[sub]{labelformat=parens}
		\begingroup
		\captionsetup{type=figure}
		\begin{subfigure}{0.3\textwidth}
			\vspace*{-5cm}
			\caption{\tt GEOLIFE}
		\end{subfigure}
		\begin{subfigure}{0.3\textwidth}
			\vspace*{-5cm}
			\caption{\tt MADRID}
		\end{subfigure}
		\begin{subfigure}{0.3\textwidth}
			\vspace*{-5cm}
			\caption{\tt BERLIN}
		\end{subfigure}
		\endgroup
	\end{minipage}
	\begin{minipage}[t]{\textwidth}
		\vspace*{-2.7cm}
		\caption{$\mathrm{OdFlowError}$ for different $M$ and $\varepsilon$ for
			each data set. The standard deviation of all 10 runs is represented
			by the error bars.}
		\label{fig:od_flows}
	\end{minipage}
\vspace{-2.5cm}
\end{figure}
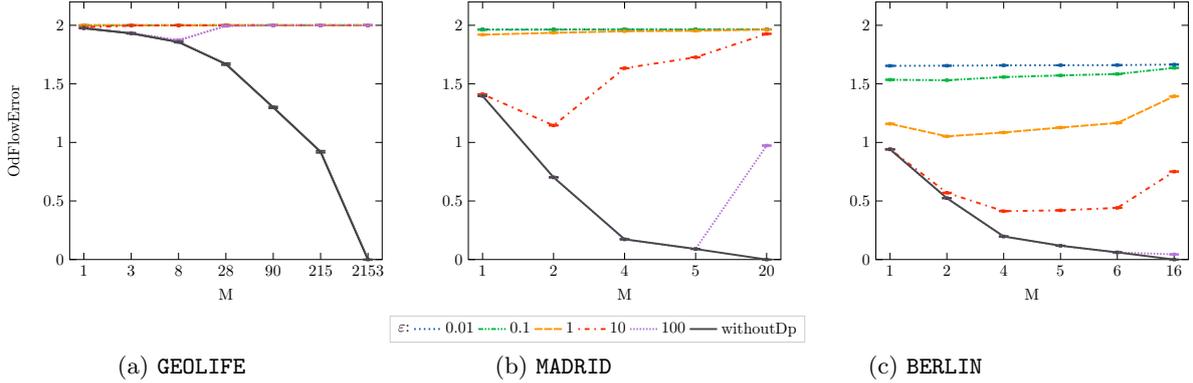

We present the results for the $\textrm{ODFlowError}$ in
Figure~\ref{fig:od_flows}. 
Recall that the 
range for the $\textrm{ODFlowError}$ lies between 0 and 2.
{\tt GEOLIFE} clearly does not contain enough data for differentially private analyses of origin-destination 
flows.
For {\tt MADRID} and {\tt BERLIN} much information is already gained with only a small $M$, as 
the error for \textit{withoutDp} shows. For $M = 4$ the $\textrm{OdFlowError}$ 
is down to 17\,\% for {\tt 
MADRID} and 19\,\% for {\tt BERLIN}. Adding noise has a major effect: Even for 
a large data set like {\tt 
BERLIN} only a privacy budget of $\varepsilon \geq 10$ falls below 100\,\%. 
These 
results raise the question 
whether 
user-level differential privacy guarantees can be given 
for OD matrices for data sets and tessellations similar in size 
to those considered here, 
while still maintaining high similarity and thereby utility of the data sets 
for further analyses.

\section{Conclusion and Lessons Learned}
\label{sec:discussion}
We have compiled typical mobility measures of human movement data to a standardized report and 
evaluated user-level privacy for selected aggregations on three practice-relevant data sets. 
Our contribution lays the groundwork for an 
easily usable open-source tool to create standardized mobility reports with differential privacy guarantees. 

In agreement with the work of Amin et al.~\cite{amin_bounding_2019} on bounding 
user contribution in the context of empirical risk minimization, 
we have showed that bounding user contribution has a major impact on error measures. 
The optimal choice of an upper bound $M$ depends on the mobility measure, data set and choice of 
privacy budget. Bounding user contribution to the 90th percentile of all users' contributions, 
and thereby downsampling all \say{power users}, has the strongest 
effect on reducing error values for most variations, unless only low guarantees are given by a comparably 
high $\varepsilon$. In this case there is, as expected, no major impact of the sensitivity, i.e. noise.

Within this work, the same random sample of size $M$ was used to compute all mobility measures. 
Future implementations should consider different sample 
sizes for different measures: E.g., absolute counts like the number of trips could be computed with 
the entire data set, while visits per location could be based on a sample with a small $M$.
The split of the privacy budget between mobility measures should further be optimized as some measures 
are 
more resistant to noise. Guidelines for meaningful choices of $\varepsilon$ and $M$ would be desirable. 
It should be noted though that the general question of what privacy budget is suitable is not
straightforward and depends on the data set and use case. 

Even for supposedly large data sets, counts of single bins shrink to comparatively small numbers as data is 
disaggregated spatially and temporally. Maintaining a high similarity to the original distribution while 
preserving privacy of individuals thus becomes a difficult task. 
Before one optimizes error values, the question should be raised how well a given data set is suited for 
user-level differential privacy. 
A data set might not include enough users for meaningful, fine-granular 
aggregations like origin-destination matrices. E.g., more than 10\,\% of all 
trips within the {\tt 
GEOLIFE} data set are produced by a single user and almost 50\,\% by 9 users. 
It is likely that these users do 
not represent 50\,\% of a population and its movements to which most use cases 
using urban mobility data 
refer. 
Next, one should be mindful about the desired analyses. A data set might be large enough for temporal and 
spatial aggregations separately, but not for combined analyses like 
visits per destination and time or OD flows. 
Further research should be dedicated to 
optimize the report's utility, e.g., by smart privacy budget splitting or eliminating queries that are not useful 
for a certain data set. Additionally, guidance on the size of a tessellation and the number of time windows 
should be provided.

Our evaluations serve as an orientation for the expected utility 
for a given privacy budget and vice versa for different mobility measures.
Further research should work towards guidelines 
that are easy to understand and apply in everyday practice.
Moreover, the compiled mobility measures 
should be assessed with practitioners to verify their usefulness. Further measures could be included, e.g., 
additional variables such as the traffic mode or user demographics. 
To increase the usability of a mobility report, margins of error resulting from adding noise will be included in 
the next version of the report.

We want to raise awareness to an additional issue that comes with user-level differential privacy: To be 
able to bound user contribution, a user identifier for each record is necessary. This contradicts the 
recommendation of data minimization to only store needed information and 
deserves
further discussions in future research.

\subsection*{Acknowledgments}
This work is part of the FreeMove project. We’d hereby thank the other project members for their 
valuable 
input.

\bibliographystyle{IEEEtran}
\bibliography{local}

\appendix

\end{document}